\documentclass[acmsmall,screen]{acmart}

\AtBeginDocument{%
  }

\usepackage{amsmath, amsthm, amsfonts}
\usepackage[ruled,vlined]{algorithm2e}
\usepackage[T1]{fontenc}
\usepackage{enumerate}
\usepackage{listings}
\usepackage{xcolor}
\usepackage{program}
\usepackage{caption}
\usepackage{subcaption}
\usepackage{graphicx}
\usepackage{algorithm2e}
\usepackage{proof}
\usepackage{tikz}
\usepackage{colortbl}
\usepackage{wrapfig}
\usepackage{stmaryrd}
\usepackage{marvosym}
\usepackage{pifont}
\usepackage[normalem]{ulem}
\usepackage{stackengine}
\usepackage{scalerel}
\usepackage{multirow}

\newcommand{\defeq}{\mathrel{ \hat{=} }} 

\usepackage{zref-savepos}
\makeatletter
\@ifundefined{zsaveposx}{\let\zsaveposx\zsavepos}{}
\makeatother
\newcounter{hposcnt}
\renewcommand*{\thehposcnt}{hpos\number\value{hposcnt}}

\makeatletter
\newcommand*{\UP}{
  \zsaveposx{\thehposcnt u}%
  \zref@refused{\thehposcnt s}%
  \zref@refused{\thehposcnt u}%
  \kern\zposx{\thehposcnt s}sp\relax
  \kern-\zposx{\thehposcnt u}sp\relax
}
\makeatother

\newcounter{hposcntimp}
\renewcommand*{\thehposcntimp}{hpos\number\value{hposcntimp}}

\makeatletter
\newcommand*{\UPimp}{
  \zsaveposx{\thehposcntimp u}%
  \zref@refused{\thehposcntimp s}%
  \zref@refused{\thehposcntimp u}%
  \kern\zposx{\thehposcntimp s}sp\relax
  \kern-\zposx{\thehposcntimp u}sp\relax
}
\makeatother

\definecolor{codegreen}{rgb}{0,0.6,0}
\definecolor{codegray}{rgb}{0.5,0.5,0.5}
\definecolor{codepurple}{rgb}{0.58,0,0.82}
\definecolor{backcolour}{rgb}{0.97,0.97,0.97}

\lstdefinestyle{pnnstyle}{
    backgroundcolor=\color{backcolour},   
    commentstyle=\color{codegreen},
    keywordstyle=\color{magenta},
    stringstyle=\color{codepurple},
    morekeywords={MAX, int1},
	basicstyle=\ttfamily\footnotesize,
    breakatwhitespace=false,
    xleftmargin=12pt,
	numbers=left,
	numbersep=4pt,
	frame=lines,
	rulecolor=\color{codepurple},
    breaklines=true,                 
    captionpos=b,                    
    keepspaces=true,
    showspaces=false,                
    showstringspaces=false,
    showtabs=false,                  
    tabsize=2,
	mathescape=true,
	framexleftmargin=0mm,
	escapechar=|
}

\lstdefinelanguage{customC}
{
language=C,
morekeywords={elsif, MAX, int1, procedure, match, with, let, in, foreach, forall, loop, MayReach},
}

\newcommand{\aps}{\textsf{AP}}

\newcommand{\sem}[1]{[\![ #1 ]\!]}
\newenvironment{itemize*}%
  {\begin{itemize}%
}
  {\end{itemize}}
\newenvironment{enumerate*}%
  {\begin{enumerate}%
}
  {\end{enumerate}}
 
\newcommand\textts[1]{\texttt{\small #1}}
  \newcommand\ignore[1]{}

\newcommand\added[1]{\textcolor{blue}{#1}}

\newcommand{\dig}{\textsf{DIG}}

\newcommand\eg{{\it e.g.,}}
\newcommand\ie{{\it i.e.,}}

\newcommand\Cond{\mathcal{C}}

\newcommand\states{\Sigma}
\newcommand\tstate[1]{\sigma_{#1}}

\newcommand\ttskip{\texttt{skip}}

\newcommand\trace{\pi}

\newcommand\prTraces{{\textsf{traces}}}

\newcommand\pc{\textsf{pc}}

\newcommand\vars{\textsf{Vars}}
\newcommand\vals{\textsf{Vals}}
\newcommand\bools{\mathbb{B}}

\newcommand\prop{\varphi}

\newcommand\FF{\textsf{F}}
\newcommand\EE{\textsf{E}}
\newcommand\A{\textsf{A}}
\newcommand\AF{\textsf{AF}}

\newcommand\AG{\textsf{AG}}
\newcommand\EF{\textsf{EF}}

\newcommand\EG{\textsf{EG}}
\newcommand\WW{\textsf{W}}
\newcommand\false{\textsf{false}}
\newcommand\stmt{\mathsf{stmt}}
\newcommand\bstmt{\mathsf{b}}
\newcommand\estmt{\mathsf{e}}

\newcommand{\hlbox}[1]{%
  \bgroup
  \expandafter\def\csname sout\space\endcsname{\bgroup \ULdepth =-.8ex \ULset}%
  \markoverwith{\textcolor{#1}{\rule[-.5ex]{.1pt}{2.5ex}}}%
  \ULon}

\newcommand\ttt[1]{\texttt{#1}}
\newcommand\ttbreak{\texttt{break}}
\newcommand\ttif{\texttt{if}}
\newcommand\ttthen{\texttt{then}}
\newcommand\ttelse{\texttt{else}}

\newcommand\ttwhile{\texttt{while}}
\newcommand\ttand{\texttt{and}}
\newcommand\ttor{\texttt{or}}
\newcommand\tttrue{\texttt{true}}
\newcommand\ttfalse{\texttt{false}}
\newcommand\ttsemi{\texttt{;}}
\newcommand\dyRefine{\textrm{\sc dyRefine}}
\newcommand\dyGeneralize{\textrm{\sc dyGeneralize}}

\newcommand\snap[1]{\mathbf{snap}^{#1}}
\newcommand\loc{\ell}
\newcommand\locs{\mathcal{L}}
\newcommand\bloc{b^{\loc}}

\newcommand\synthposb{{b}_{pos}}
\newcommand\synthnegb{{b}_{neg}}
\newcommand\synthposbi{{b}^{\added{\loc}}_{pos}}
\newcommand\synthnegbi{{b}^{\added{\loc}}_{neg}}

\newcommand{\convexhullOR}{\stackMath\mathbin{\stackinset{c}{0.05ex}{c}{-0.1ex}{\vee}{\bigcirc}}}

\newcommand\Tool{\textsf{DrNLA}}
\newcommand\tool{\Tool}
\newcommand\FuncTion{{\sc FuncTion}}

\newcommand\rUNK{??}
\newcommand\rTRUE{$\checkmark$}
\newcommand\rFALSE{{\bf X}}

\newcommand\TrimPos{\textsf{TrimPos}}
\newcommand\ExpandNeg{\textsf{ExpandNeg}}
\newcommand\TrimNeg{\textsf{TrimNeg}}
\newcommand\ExpandPos{\textsf{ExpandPos}}

\newcommand\cexpath{\mathit{cex}}

\newcommand\benchsvcomp{\texttt{ctlnla-oopsla20}}
\newcommand\benchpldi{\texttt{ctlnla-pldi13}}
\newcommand\benchcustom{\texttt{ctlnla-custom}}

\newcommand\preds[1]{\textsf{preds}(#1)}

\newcommand{\errorPosTooB}{\texttt{err}_{\textsf{TrimPos}}}
\newcommand{\errorPosTooS}{\texttt{err}_{\textsf{ExpandPos}}}
\newcommand{\errorNegTooB}{\texttt{err}_{\textsf{TrimNeg}}}
\newcommand{\errorNegTooS}{\texttt{err}_{\textsf{ExpandNeg}}}

\newcommand\Pcheck{P_\textsf{valid}}

\newcommand\Psnap{P_\textsf{snap}}

\acmVolume{1}
\acmNumber{ARXIV} 
\acmYear{2023}
\acmMonth{1}
\acmDOI{} 
\startPage{1}

\setcopyright{none}

\usepackage{booktabs}   
\usepackage{subcaption} 
\begin{document}

\title{DrNLA: Extending Verification to Non-linear Programs through Dual Re-writing}



\author{Yuandong Cyrus Liu}
\affiliation{\institution{Stevens Institute of Technology} \country{USA}}
\author{Ton-Chanh Le}
\affiliation{\institution{Stevens Institute of Technology} \country{USA}}
\author{Timos Antonopoulos}
\affiliation{\institution{Yale University} \country{USA}}
\author{Eric Koskinen}
\affiliation{\institution{Stevens Institute of Technology} \country{USA}}
\author{ThanhVu Nguyen}
\affiliation{\institution{George Mason University} \country{USA}}

\begin{abstract}

For many decades, advances in static verification have focused on linear integer arithmetic (LIA) programs. Many real-world programs are, however, written with non-linear integer arithmetic (NLA) expressions, such as programs that model physical events, control systems, or nonlinear activation functions in neural networks. While there are some approaches to reasoning about such NLA programs, still many verification tools fall short when trying to analyze them.

To expand the scope of existing tools, we introduce a new method of converting programs with NLA expressions into semantically equivalent LIA programs via a technique we call \emph{dual rewriting}. Dual rewriting discovers a linear replacement for an NLA Boolean expression (e.g. as found in conditional branching), simultaneously exploring both the positive and negative side of the condition, and using a combination of static validation and dynamic generalization of counterexamples. While perhaps surprising at first, this is often possible because the truth value of a Boolean NLA expression can be characterized in terms of a Boolean combination of linearly-described regions/intervals where the expression is true and those where it is false.

The upshot is that rewriting NLA expressions to LIA expressions beforehand enables off-the-shelf LIA tools to be applied to the wider class of NLA programs. We built a new tool \Tool{} and show it can discover LIA replacements for a variety of NLA programs. We then applied our work to branching-time verification of NLA programs, creating the first set of such benchmarks (92 in total) and showing that \Tool{}'s rewriting enable tools such as \FuncTion{} and T2 to verify CTL properties of 42 programs that previously could not be verified. We also show a potential use of \Tool{} assisting Frama-C in program slicing, and report that execution speed is not impacted much by rewriting.
\keywords{Dynamic Analysis \and Dynamic Verification \and Program Simplification}
\end{abstract}
\maketitle


\section{Introduction}

Rapid progress in advanced SMT solving has revolutionized the automation of reasoning in real-world software scenarios. For instance, Amazon performs billions of SMT queries every day using this technology to ensure the correctness and reliability of AWS cloud computing~\cite{Rungta2022}. Among many theories supported by modern SMT solving, linear arithmetic (LIA), which is commonly used in real-world software, is well-supported, has many advanced LP optimization techniques, and is the standard in annual competitions that benchmark SMT solving~\cite{WCDHNR19}. In addition, verifying and reasoning about programs with LIA is a major focus of numerous verification tools, \eg~both the T2~\cite{T2} and \FuncTion{}~\cite{function} tools for analyzing temporal properties are mainly designed for LIA programs and thus work well on them.

However, with the increasing complexity of our world and the advancements in automated reasoning, researchers are now presented with opportunities to tackle more challenging problems that involve \emph{nonlinear operations}. These operations appear in numerous scientific, engineering, safety, and security-critical applications. However, compared to LIA, nonlinear arithmetic (NLA) has much less support in constraint solving and modern software analysis tools. In many cases, modern SMT solvers return unknown or time out when dealing with formulas involving nontrivial NLA expressions. 
Temporal verification tools (\eg~T2 and \FuncTion{}) similarly struggle when confronted with NLA programs. We found that these programs are just not geared for NLA programs: \FuncTion{} reported unknown and T2 emitted meaningless output (\eg~``valid'' even on invalid inputs).

\smallskip
In this paper we take a new approach to programs with NLA expressions: translating them to \emph{semantically equivalent} programs in which NLA expressions have been replaced with LIA expressions. 
Consequently, one can then leverage the power of existing static analysis and verification tools on those transformed programs. 
To this end, we introduce an integrated mix of static and dynamic analysis in what we call \emph{dual rewriting}. We analyze a given Boolean NLA expression $b_{NLA}$ in a branch condition or loop guard and synthesize LIA expressions that are equivalent replacements. 
The replacement expressions are boolean combinations of linear equalities/inequalities, thus trading off NLA complexity for boolean LIA complexity.
Surprising as this may sound, it is often possible because the truth value of a Boolean NLA expression (for example single-variable NLA expression $x^2>25$) can be characterized in terms of the regions/intervals where the expression is true and those where it is false. Those regions can be described as Boolean combinations of \emph{linear} expressions (\eg~$x < 5 \vee x > 5$).
Dual rewriting also exploits the \emph{program context} of a $b_{NLA}$ in the source program, so that a replacement $b_{LIA}$ need not be a closed-form equivalent of $b_{NLA}$, but rather only equivalent in the specific program context.

\paragraph{Challenges and Contributions}
The main benefit of dual rewriting is that it can be used as a pre-processing phase for existing static verification or analysis tools, allowing those tools to be applied to broader settings. We now summarize how we address challenges in this direction.

\emph{1. Refinement algorithm.} (\S\ref{sec:refinement}) \emph{How can dual rewriting be formulated?}
We describe the dual rewriting algorithm to analyze Boolean NLA expressions and iteratively synthesize equivalent Boolean LIA expressions. The algorithm follows a somewhat typical refinement loop structure. However, 
rather than synthesizing a single $b_{LIA}$, we instead simultaneously synthesizes LIA expressions for both the positive side of a given NLA Boolean expression $b_{NLA}$, denoted $\synthposb$, and the negative side of $b_{NLA}$ (\ie~$\neg b_{NLA}$), denoted $\synthnegb$.
We found this was essential to making the algorithm effective as it could just as quickly rule ranges out as it ruled other ranges in. Consequently the refinement algorithm has four different counterexample situations, depending on whether $\synthposb$ (resp, $\synthnegb$) over- or under-approximates the positive side (resp, negative side) of $b_{NLA}$.
Our algorithm is sound in that it emits LIA expressions that are \emph{equivalent} to their NLA counterparts in the respective program contexts at hand and, later in \S\ref{sec:termination}, we discuss various conditions under which it terminates.

\emph{2. Context-sensitive static validation.} (\S\ref{sec:validate})
\emph{Must we synthesize LIA expressions that are equivalent to their NLA counterparts in all contexts?}
We show that this is not necessary and give a novel context-sensitive static validation step in the refinement algorithm, causing the overall algorithm to target a context-sensitive output.
We give a program transformation $\Pcheck{}$ that checks context sensitive equivalence of $b_{NLA}$ expresions with candidate $\synthposb/\synthnegb$ pairs in the program context where $b_{NLA}$ occurs. To this end, it constructs the four-way check as to whether $\synthposb/\synthnegb$ under-/over-approximate their NLA counterpart in that program context.

An important consequence is that
the transformed program $\Pcheck{}$ is a reachability program \emph{that actually can often be addressed with existing solvers} such as Ultimate~\cite{ultimate}, despite the fact that 
validating $\Pcheck{}$ requires some degree of NLA reasoning.
However, the problem is now significantly simpler because
the solver is merely checking equivalence with a \emph{provided} candidate LIA expression, rather than trying to synthesize one from scratch. Furthermore, $\Pcheck{}$ is a safety problem and reasoning for more sophisticated properties (\eg~termination, non-termination, temporal logic, etc.) is not necessary.
Consequently, we can often exploit these existing solver's NLA support for safety validation as a building block.

\emph{3. Dynamic counterexample generalization.} (\S\ref{sec:dygeneralize})
\emph{How can refinement be practical with only single-state counterexamples?}
We show that the refinement loop can be expedited through a novel dynamic form of generalizing counterexamples.

While static validation counterexamples offer a data point to refine $\synthposb$ or $\synthnegb$,
refining based on merely a single data point does not lead to an algorithm that converges. 
We thus introduce 
a static/dynamic method for generalizing a single counterexample.
We employ an SMT solver to generate many states that all
witness shortcomings of $\synthposb$ or $\synthnegb$,
and then we use dynamic analysis to learn linear conditions over these states.
These linear arithmetic expressions are then conjunctively or disjunctively combined with the 
accumulating $\synthposb/\synthnegb$ pair.

\emph{4. Implementation: \Tool.} (\S\ref{sec:impl})
\emph{Can dual rewriting be realized in an implementation?}
We implemented the rewriting technique in a new tool \tool{} that analyzes programs and emits 
a mapping from the NLA expressions in the program to equivalent LIA replacements.
\tool{} is written in Python and OCaml and built on top of CIL, Z3, Ultimate (for static validation), and DIG~\cite{tosem2013} (for dynamic learning).  
Our overall dual rewriting algorithm also employs unsat core and convex hull techniques to further improve performance.
We plan to make \Tool{} open source and available on GitHub. The sources of \Tool{} and all benchmarks are available at the FigShare service: \url{https://figshare.com/s/a4df43da932fd2d95fcb}.

\emph{5. Evaluation.} (\S\ref{sec:eval}) \emph{How does \tool{} perform?} Since there are many possible applications of dual rewriting, we chose one salient one: branching-time verification of 
Computation Tree Logic (CTL)~\cite{Clarke1986} properties; despite many existing techniques and tools~\cite{Cook2011,CookK2013,Beyene2013,Cook2017,Tellez2017,Urban2018}, there are currently no tools capable of verifying CTL properties of NLA programs.

\added{\emph{(5.i)}} Since no NLA+CTL programs exist, we first construct \textbf{92 benchmarks} with NLA programs with CTL properties (\S\ref{sec:benchmarks}). The state-of the-art CTL verifiers \FuncTion{} and T2 
cannot handle these benchmarks: \FuncTion{} returns ``unknown'' for every program and T2 is apparently not meant to accept them as input (it marks most invalid benchmarks as valid).
We will contribute these benchmarks (and our \tool{} implementation) in open source format.

\added{\emph{(5.ii)}} We next evaluate whether \tool{} can discover LIA expressions and found that \Tool{} can synthesize LIA replacements for \textbf{86/92} benchmarks (\S\ref{sec:results}). 
Typically these replacements can be found within a few minutes (due to the need to execute the program), although in future work this could likely be reduced by partially re-using executions across refinement iterations.
Although our goal is enabling analysis rather than actually executing the transformed program, we nonetheless confirmed that the execution speed of the transformed program is comparable to that of the original.

\added{\emph{(5.iii)}} We then show that, with the help of \tool{},  off-the-shelf CTL tools T2 and \FuncTion{}, for the first time, can be applied to verify CTL properties of NLA programs (\S\ref{sec:improvements}). Specifically, these tools (especially T2), when applied to the the transformed programs, were able to prove or disprove \textbf{42/92} programs. 
Given that no modifications were needed to these off-the-shelf tools, the few minutes needed to pre-process with \tool{} seems justified.

\added{\emph{(5.iv)}}
Beyond CTL temporal verification there are other analysis/verification domains where dual rewriting can be applied, which we plan to investigate in future work. As an indicator of this hypothesis, we report a small example where dual rewriting can enable program slicing plugin in Frama-C~\cite{baudin2021dogged} to be applied to an NLA program, where previously it could not (\S\ref{sec:slicing}).

\paragraph{Other related works.}
The well-known Ultimate~\cite{ultimate} tool has support for some non-linear expressions for proving reachability, termination or linear temporal logic (LTL). 
Ultimate does not support branching-time logics and instead exploits trace techniques to verify LTL properties, which could not be used for CTL. Specifically, 
since LTL is a trace-based logic, traces can be taken one path at a time and, for each case, over-approximated. 
Such trace-by-trace over-approximation does not work well for branching-time, where an over-approximation of a \texttt{then} branch is an under-approximation of the corresponding \texttt{else} branch.
Dynamic invariant inference works such as ~\cite{yao2020learning,nguyen2021using} can find NLA invariants
or rank functions~\cite{le2020dynamite,uterm2022}, but do not support branching-time properties such as CTL.
Other works~\cite{characterizing_2014, jobstmann_2016, artho_polynomial_2016, hrushovski_polynomial_2018} focus on generating polynomial invariants and recently, authors of ~\cite{unsolvable-loops22} presented a method for so-called unsolvable loops in the setting of recurrence relations.
Still others~\cite{porous_2021} introduce porous invariants that are not necessarily linear.



\section{Overview}\label{sec:overview}

Consider the program to the right, adapted from \texttt{cohencu.c} in the SV-COMP non-linear benchmarks\footnote{\url{https://gitlab.com/sosy-lab/benchmarking/sv-benchmarks/-/blob/main/c/nla-digbench/cohencu.c}}. 
Although this program seems simple, it has a polynomial loop guard, making it difficult
\begin{wrapfigure}[8]{r}[0pt]{2.8in}
\begin{lstlisting}[language=Python,style=pnnstyle]
int y=1, z=6, c=0, p=2;   
int k=*; 
while (z*z - 12y - 6z + 12 + c <= k):
    y = y + z;
    z = z + 6;
    c = c + 1;
    p = 1;
\end{lstlisting}
\end{wrapfigure}
for tools to discover and validate loop invariants.
The idea we explore in this paper is to develop an algorithm that is able to automatically discover a semantically equivalent replacement for the polynomial \lstinline|z*z - 12y - 6z + 12 + c <= k| that is instead expressed in (perhaps a boolean combination of) linear expressions.
Consequently, more existing static tools could be applied on the re-written program. In this section we will describe our work, using the example to the right.

\paragraph{Learning linear conditions}
Many static verification tools struggle with NLA reasoning and, thus, cannot be directly used to analyze them. However, a variety of works in recent years have shown that dynamic analysis can learn linear behaviors of non-linear programs by analyzing concrete executions and inferring correlations. This has been done for invariants~\cite{tosem2013,nguyen2021using,nguyen2012using,yao2020learning}, termination~\cite{le2020dynamite,uterm2022},
separation logic~\cite{chanhSeparationLogic}, etc. 
Many such works combine dynamic learning to find candidate invariants or ranking functions, with static validation to check sufficiency. Although this means that static tools are used to do some nonlinear reasoning, it is only for the purpose of validation, not for the harder problem of inference and search.
We exploit this general strategy in this paper, but aim now at the specific problem of synthesizing equivalent LIA alternatives to NLA conditions.

The key idea of our procedure is to identify NLA Boolean conditions (and challenging linear conditions) and for each such condition $b$, to simultaneously synthesize 
two linear conditions $\synthposb$ and $\synthnegb$, the former reflecting the conditions under which $b$ holds and the latter reflecting the conditions under which $\neg b$ holds.
Our algorithm aims for these conditions to be \emph{exact} (\ie~these conditions will neither be an over-approximation nor an under-approximation), since branching-time verification requires that we reason precisely about branching.
With exact alternatives we can replace the NLA condition with an LIA condition, without changing the semantics of the program. Further more, accumulating these conditions \emph{together} enables us to explore from both directions  until we have  exactly captured both truth values of $b$. 

We now describe dual rewriting, using the above example. In this case,
$b$ is the loop guard
\ttt{z*z - 12y - 6z + 12 + c <= k} and naturally $\neg b$ is its negation.

{\bf Step 1.} Initial guess for $\synthposb$ and $\synthnegb$. This can be done by simply executing the program on random inputs and 
instrumenting the program to capture the values of variables
\ttt{y,z,c,k} inside the loop and, separately, capture the values of those variables after the loop. The first set of values are examples where $b$ holds and the second set is where $\neg b$ holds. Using off-the-shelf learning procedures~\cite{tosem2013,nguyen2021using} we can infer  candidate invariants for these two program locations
representing our initial guesses for 
$\synthposb$ and $\synthnegb$, with the caveat that they are only sound for the random inputs considered so far. 
After instrumenting the program as described above, the DIG tool~\cite{tosem2013,nguyen2021using} learns many possible candidate invariants as follows:
\[\begin{array}{ll}
\text{For } \synthposb :&  \{ 2 \ge p, -p \le -1, 0 = -6c + z - 6, -p - z \le -8, 0 \ge -c, 0 \ge c - k\} \\
\text{For } \synthnegb  : & \{0 \ge -c + p, 0 \ge -k, 0 = -6k + z - 12, 0 = p - 1, 0 = c - k - 1\}
\end{array}\]
Due to the nature of dynamic analysis which derives conditions that hold at the program locations of interest w.r.t just the snapshots at those locations, the above guesses for $\synthposb$ and $\synthnegb$ contain many irrelevant conditions to the expected condition $b$ and $\neg b$. For example, they have loop invariants at the locations (such as $0 = -6c + z - 6$ in $\synthposb$) or conditions specific to the given snapshots (such as $0 = -6k + z - 12$ in $\synthnegb$). However, $\synthposb$ and $\synthnegb$ also have conditions like $0 \ge c - k$ and $0 = c - k - 1$ which are close to the desired result $b \equiv c \le k$ and $\neg b \equiv c > k$, respectively. Since the condition $b$ and its negation $\neg b$ always contradict each other, we  only keep conditions in the guesses $\synthposb$ and $\synthnegb$ which also contradict the conditions in the other group. Such conditions can be found via the unsatisfiable core of $\bigwedge(\synthposb \cup \synthnegb)$, which are the set of conjuncts $\{0 \ge c - k, 0 = c - k - 1\}$. We then only use conditions in $\synthposb$ and $\synthnegb$ which are also in the unsatisfiable core. This leads us to the first
guess pair:
\[
\text{First Iteration: } \;\; 
\synthposb \equiv 0 \geq c - k
\text{ and }
\synthnegb  \equiv 0 = c - k - 1\]

{\bf Step 2.} Validating the pair $(\synthposb,\synthnegb)$.
The next step is to validate whether 
$\synthposb \Leftrightarrow b$ and 
$\synthnegb \Leftrightarrow \neg b$. If we got lucky and it holds we are done. Otherwise, there are four possible cases to consider as \begin{wrapfigure}[8]{r}[34pt]{3.0in}
\includegraphics[width=2.5in]{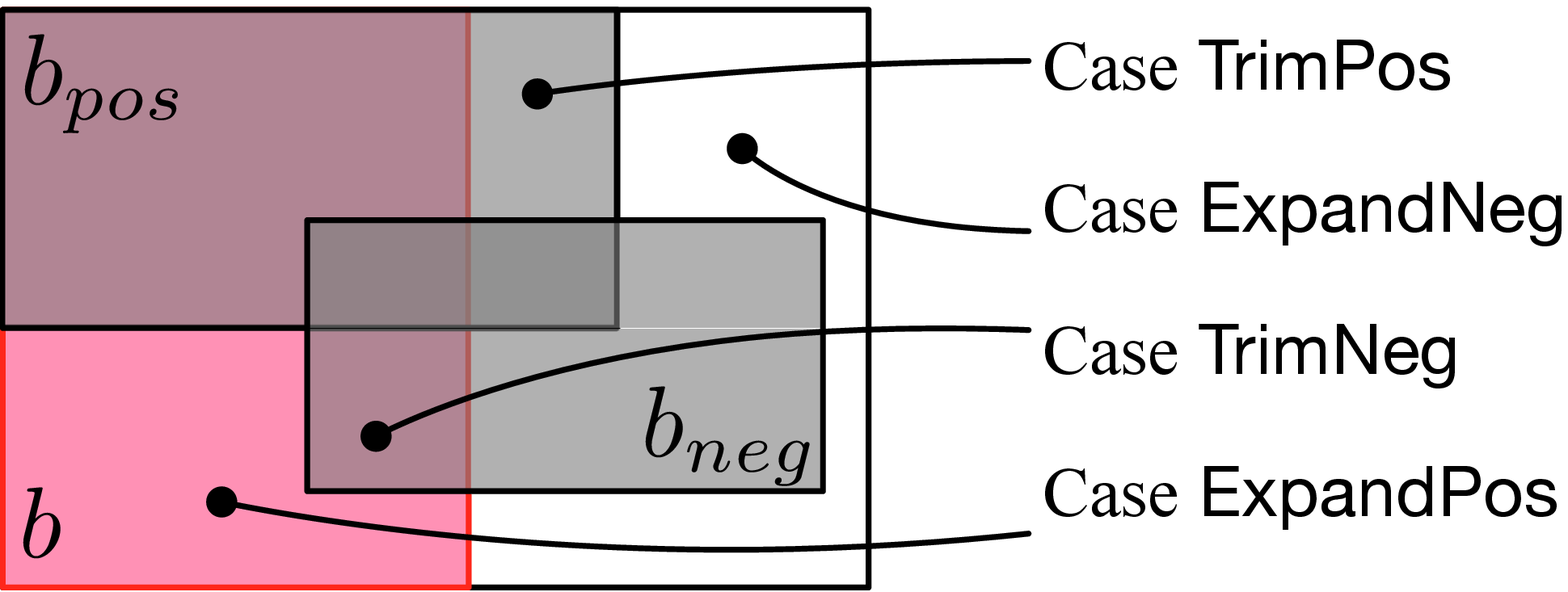}
\end{wrapfigure}
depicted in the diagram to the right.
The pink area reflects the goal condition to match. Our guess for the positive side (grey box $\synthposb$) could be incorrect in two possible ways: 
In case \TrimPos{} it includes executions where $\neg b$ holds and must be trimmed down, and 
in case \ExpandPos{} it does not include all executions where $b$ does hold.
Dually, our guess for the negative side (grey box $\synthnegb$) could be incorrect in two possible ways: 
In case \TrimNeg{} it includes executions where $b$ holds and must be trimmed down, and 
in case \ExpandNeg{} it does not include all executions where $\neg b$ does hold.

\begin{figure}
\centering
\includegraphics[width=4.5in]{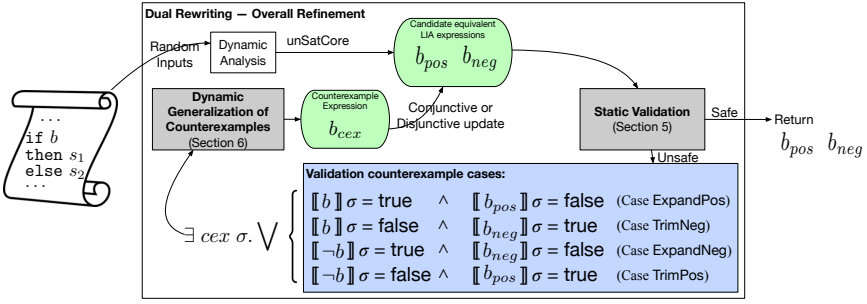}
\caption{\label{fig:ddr} Overall flow of the Dual Rewriting algorithm.}
\end{figure}

How do we know which case applies to the current $(\synthposb,\synthnegb)$ pair?
In \S\ref{sec:validate} we describe a method for static validation through a program transformation that maintains the context in which $b$ occurs and emits counterexamples that indicate which trim/expand positive/negative case needs to be refined next.
When the validation returns ``safe'', 
$\synthposb$ is equivalent to $b$, $\synthnegb$ is equivalent to $\neg b$ and, naturally,
$\synthnegb$ is the negation of $\synthposb$.
When attempting to validate our first guess $\synthposb,\synthnegb$ for the running example, we may find a counterexample consisting of the following valuation of variables $c=0, k=-2, p=2, y=1, z=6$.
For these valuations,
the original negated condition
$\neg b \equiv \ttt{z*z - 12y - 6z + 12 + c > k}$ is true
because $36-12-36+12+0>-2$,
yet our approximation of this negated condition
$\synthnegb \equiv \ttt{0 = c - k - 1}$ is false because $0\neq 1$.
This case is \ExpandNeg: we must increase the size of $\synthnegb$ so that it is true for this state.
The counterexamples returned by \S\ref{sec:validate} are next used to refine $\synthposb$ and $\synthnegb$, as we will discuss in a moment. 
The overall algorithm is depicted in Fig.~\ref{fig:ddr}.
So far we discussed how the input program progresses to the initial guess (in green) and how static validation yields a state $\sigma$ that is one of four possible counterexamples (in blue). Below we will discuss the next gray box, used to generate a new expression $b_{cex}$ that can conjunctively/disjunctively amend $\synthposb$ or $\synthnegb$, as the case may be.

{\bf Step 3.} Dynamic Generalization of Counterexamples.
A single-state counterexample is not enough to enable a refinement procedure to be tractable.
In \S\ref{sec:dygeneralize} we discuss a dynamic analysis procedure that goes beyond a single-state model of a counterexample and instead generates many models and then learns an expression $b_{cex}$ from them. In this way, $b_{cex}$  captures more of the way in which
$\synthposb$ or $\synthnegb$ must be amended than would be by a single state.
For example, the validation for the program above failed and found that 
$\synthnegb\equiv \ttt{0=c-k-1}$. While this single condition could be helpful it is rather a restrictive case being an equality.
Our procedure instead generates many models of the counterexample and then
uses \dig\ to learn an expression capturing those models.
After some filtration via the UNSATcore, we obtain $b_{cex} \equiv 0 \geq p + k \wedge  0 = p - 2 \wedge 0 = c$.

We next need to disjunctively combine this $b_{cex}$ with $\synthnegb$ (because we were in the \ExpandNeg{} case). However, a direct logical disjunction $\synthnegb \vee b_{cex}$ does not always work well because $b_{cex}$ is still a bit too concrete and specific to this counterexample. By using the convex hull, however, we can generalize the current $\synthnegb$ to be weak enough to also include $b_{cex}$.
We therefore attempt to compute the convex hull~\cite{barber1996quickhull} of the disjunction which, in this case, yields
$0 + k - c \leq -1$. 
We now have our second guess:
\[\text{Second Iteration: }\;\; \synthposb \equiv 0 \geq c - k \text{ and }
\synthnegb  \equiv k - c \leq -1\]
With this, we return to the validation phase and, at this point, validation does not find any counterexamples and indeed proves that these conditions are equivalent to the NLA expression in the loop guard of the original program.
Therefore, it is sound to replace the original program's polynomial loop guard with
$c-k \leq 0$, and maintain the program's behavior.

\emph{The \Tool{} Tool.}
We implemented the above techniques in a new tool called \Tool{}. We discuss its implementation in \S\ref{sec:impl} and evaluate it in \S\ref{sec:eval}. The upshot is that \Tool{} can be used as a pre-processing step to transform an NLA program into an LIA program, which can then be analyzed by existing tools in domains such as temporal verification that cannot handle NLA natively.

\subsection{Application: Branching-time Verification}

With the advent of dual rewriting, we next consider the important question: where can dual rewriting be useful? To that end, we consider the example application of branching-time verification.
Numerous techniques~\cite{Cook2011,CookK2013,Beyene2013,Cook2017,Tellez2017,Urban2018}
and tools such as T2~\cite{Cook2015} and \FuncTion~\cite{function} 
have been developed for statically verifying CTL properties of programs.
Such state-of-the-art CTL verification tools perform well on programs and properties involving linear integer arithmetic (LIA), \eg~programs with LIA expressions in assignment statements and branch conditions.
Though these tools perform well on LIA programs and properties, they have limited support for programs with non-linear arithmetics (NLA). In our experiments, the CTL verification tool \FuncTion, for example, would return `Unknown' on  programs with NLA properties, even simple ones such as $x^2 > 49$.  T2 also does not appear to apply to NLA programs. (T2 actually `Valid' on most examples, even on invalid examples.)

\begin{figure}[t]
\begin{tabular}{cc}
\begin{minipage}[t]{0.48\textwidth}
\begin{lstlisting}[language=Python,style=pnnstyle]
int y=1, z=6, c=0, p=2;   
int k=*; 
while (z*z - 12y - 6z + 12 + c <= k):
    y = y + z;
    z = z + 6;
    c = c + 1;
    p = 1;
p = 0; |\label{ln:ex1p}|
return 0;
\end{lstlisting}
\centering{Valid: $\EF(p=0) \wedge \EF(p=1)$} 
\end{minipage}
&
\begin{minipage}[t]{0.48\textwidth}
\begin{lstlisting}[language=Python,style=pnnstyle]
int y=1, z=6, c=0, p=2;   
int k=*;   
while (z*z - 12y - 6z + 12 + c <= k):
    y = y + z;
    z = z + 6;
    c = c + 1;
    p = 1;
p = c-k;
return 0;
\end{lstlisting}
\centering{Invalid: $\EF(p=0) \wedge \EF(p=1)$}
\end{minipage}
\end{tabular}
\caption{\label{fig:ex1} Running examples of programs with NLA expressions and CTL properties.}
\end{figure}

\newcommand\pp{\ttt{p}}
\newcommand\cc{\ttt{c}}
\newcommand\kk{\ttt{k}}

Let us consider an example NLA program and CTL property  shown in Fig.~\ref{fig:ex1}. Here we have extended the earlier SVCOMP \texttt{cohencu.c} example, creating two CTL benchmarks.
Here we have added atomic propositions over the variable $\pp$ and a CTL property $\EF(p=0) \wedge \EF(p=1)$. The CTL property states that from the start state
there is at least one execution that leads to a state where $\pp=0$ holds and a state where $\pp=1$ holds.

Non-linear reasoning is needed to validate/invalidate these examples.
Assuming precondition \ttt{c <= k} holds initially, the CTL property holds for the left program. Temporal verification requires non-linear reasoning to determine that the loop terminates, thus leading to the $\pp=0$ state, as well as to determine the feasibility of at least one loop iteration in order to reach the $\pp=1$ state. 
On the other hand, the CTL property does not hold for the right program. To invalidate the property, verification also requires non-linear reasoning to discover the loop invariant \ttt{z*z - 12*y - 6*z + 12 == 0} (thus the loop condition is equivalent to \ttt{c <= k}) and determine that the assignment of \ttt{c - k} into \ttt{p} at Line~\ref{ln:ex1p} is reachable only when \ttt{c > k} (\ie~after the loop terminates). As a result, \ttt{p} is always greater than 0, thus the conjunct $\EF(\pp=0)$ is not valid from the initial state and thus the overall property does not hold. 

This requisite NLA loop condition reasoning is not available in existing CTL verification tools.
For example, the T2~\cite{T2} verifier 
accepts these kinds of programs and properties as input but likely should be rejecting them. (It clearly does not support them, as it reports Valid on most invalid examples.)
Meanwhile, the \FuncTion{} abstract interpreter~\cite{function} promptly reports ``unknown'' for both of examples.

However, if one first applies \Tool{} to re-write these programs in a pre-processing step, then these CTL verification tools can then be applied, thus expanding the scope of these tools to NLA programs.
In this example the loop guard can be rewritten to \lstinline|c-k <= 0| (the LIA expression discovered above), and then this example can be verified by T2 (and other rewritten programs can be verified by \FuncTion{}.
More generally, we report our experimental results in \S\ref{sec:improvements}, showing that \Tool{} enables \FuncTion{} and T2 to verify 42 NLA programs that previously could not be verified.
Since no NLA CTL benchmarks exist we adapted 
the DynamiTe NLA termination/nontermination benchmarks~\cite{le2020dynamite} to have CTL properties
(\benchsvcomp{}), we adapted other known 
CTL benchmarks~\cite{CookK2013} to include NLA expressions (\benchpldi{}),
and we created a set of handcrafted examples that further demonstrate the refinement process (\benchcustom{}).
\begin{wrapfigure}[10]{r}{2.8in}\centering
\begin{minipage}[t]{2.8in}
\begin{lstlisting}[language=Python,style=pnnstyle]
int a=0, s=1, t=1, k=*, c=0, p=0, x=5;
while (t*t - 4*s + 2*t + 1 + c <= k):
  a = a + 1; t = t + 2;
  s = s + t; c = c + 1;
while (x >= 0):
  if (*) x--;
p = 1;
\end{lstlisting}
\end{minipage}\\
Property: $\AF(\EF(\ttt{p}>0))$
\end{wrapfigure}
To the right is another example program from \benchsvcomp{}.
This program involves an NLA loop on Line 2. The CTL property says that across all paths, eventually a state is reached, from which point, there is at least one path to reach a state where \ttt{p} is positive.
\Tool{} enables T2 to validate this program/property, when previously it could not.

\subsection{Preliminaries}\label{sec:prelim}

\emph{Syntax.}
We work with a simple model of programs that supports nonlinear arithmetic expressions (branch/loop conditions and assignments):
\[
  \small
  \begin{array}{lllllllll}
\stmt &::=& \stmt\ \ttsemi\ \stmt 
\mid \ttif\ \bloc\ \ttthen\ \stmt\ \ttelse\ \stmt\ \mid v := e\ \mid\ \ttskip\\
& &\mid \ttwhile(\tttrue)\ \ttif(\lnot \bloc)\ \ttthen\ \ttbreak\ \ttelse\ \stmt
& \estmt &::=& c \mid v \mid \estmt \otimes \estmt \mid \ldots \\
\bstmt &::=& \tttrue\ \mid \ttfalse\ \mid \neg \bstmt \mid \bstmt\ \ttand\ \bstmt\ \mid\ \bstmt\ \ttor\ \bstmt\ \mid\ \estmt \lessapprox \estmt
& P &::=& \stmt
\end{array}\]
where $ \lessapprox ::= < \mid \leq \mid = \mid \ldots $ are comparisons between integer expressions, $\otimes$ are typical integer operations, and $c$ refers to integer constants.
We represent \ttwhile{} loops with \ttif{} and \ttbreak{}, so that we can apply our algorithms to \ttif{} statements only, yet still capture branching in loop headers.
We label branch condition expressions $\bloc$ and assume that each $\loc$ is chosen to uniquely identify the expressions, \ie~a program location from the set of locations $\locs$.
We also assume that the program location is stored in a variable \pc{} for the program counter whose value is $\loc$ when evaluating expression $\bloc$.

\emph{Semantics.} We assume a simple model of program states 
$\states : \vars \rightarrow \vals$, mapping variables to (typed) values and a standard operational semantics $R$ on statements. 
 A {\bf trace} $\trace$ of a program $P$ is a sequence $\trace = \tstate{0},\tstate{1},\ldots$ such that $\tstate{0}$ is an initial state and $\forall i \geq 0. R(\tstate{i},\tstate{i+1})$, \ie~the transitions are in the operational semantics of $P$. The set of all traces of a program $P$ is denoted by $\prTraces(P)$ and for a trace $\trace=\tstate{0},\tstate{1},\ldots$, the $j$-th state $\tstate{j}$ in the trace is denoted by $\trace[j]$.
We write the semantics of Boolean conditions  as $\sem{\bstmt} : \states \rightarrow \bools$ and expressionsas  $\sem{\estmt} : \states \rightarrow \vals$. 
We write $\preds{\bloc}$ to be the set of all reachable states under which $\bloc$ could be evaluated, \ie~$\preds{\bloc} = 
\{ \tstate{} \mid \exists \trace = \tstate{0},\ldots,\tstate{},\ldots \wedge 
\tstate{}(\pc) = \loc \}
$.

\emph{Snapshots.}
As is common in dynamic analysis, instrumentation is used so that when the program is executed, states can be recorded. We introduce a family of ``snapshot'' keywords $\snap{\loc}$ to represent statements added to the program at location $\loc$ that read and log the current state, along with the location $\loc$ where the snapshot is taken before the $\loc$ statement is executed. Apart from the side-effect of saving the state, the semantics of $\snap{}$ is otherwise \ttskip. 

We use snapshots in the following program transformation, wherein 
for each $\bloc$, one snapshot $\snap{\loc,pos}$ is added immediately inside the positive `then' branch and 
$\snap{\loc,neg}$  immediately inside the negative `else' branch:
\begin{definition} For program $P$, instrumentation for snapshots is via the following transformation:
$$
 \Psnap{}[\loc] \defeq
 \left[
   \ttif\ \bloc\ \ttthen\ s\ \ttelse\ s'\ 
 \leadsto \left(
 \begin{array}{lll}
 \ttif\ \bloc\ \ttthen\ \snap{\loc,pos}; s\ \\
 \ttelse\ \snap{\loc,neg}; s'
 \end{array}\right) \right]
$$
\end{definition}
%
%

Once a program has been transformed and then executed, the union of the states collected from these snapshot sets is a subset of $\preds{\bloc}$, but discriminates between the states where $\bloc$ holds and where $\neg \bloc$ holds.

\emph{Approximating sampled states.}
For a set of states $S$, we define an \emph{over-approximation of a sampling}, denoted $\alpha_S$, to be an abstraction that is guaranteed to be a sound abstraction only of the states in $S$. If $S\supseteq\states$, then $\alpha_S$ is sound for the program.
We assume a facility for learning abstractions $\alpha_S$ of a sample of states $S$. Such techniques are available from recent tools such as DIG~\cite{tosem2013}, Diakon~\cite{ernst2007daikon}, etc.~and provide the following function:

\begin{definition}[Learning sample approximations]
For a set of states $S\subseteq\states$, we assume the availability of a function $\textsf{learn} : S \rightarrow \bar{\Cond}$ such that $\forall \tstate{} \in S, c\in\bar{\Cond}$ we have that $\sem{c}\tstate{}$ is true.
\end{definition}

\emph{Static verification.}
We assume special locations denoted \ttt{err} and the static verification problem is to determine whether \ttt{err} is reachable in a given program.
If so, static validation returns a counterexample \emph{path} $\cexpath$. We assume that this counterexample includes the error location that was possibly reachable.

\emph{Computation Tree Logic (CTL).}
We will discuss CTL as an application in \S\ref{sec:improvements}, but here briefly recall the syntax.
The standard semantics of formulae are shown in Apx.~\ref{apx:ctlsem}.
A CTL formula is given by the following grammar:
\[
\prop ::= 
p \in \aps \mid
\prop \vee \prop \mid 
\prop \wedge \prop \mid
\AF \prop \mid
\EF \prop \mid
\A \lbrack \prop \WW \prop \rbrack \mid
\EE \lbrack \prop \WW \prop \rbrack \mid
\]
Intuitively, $\AF p$ means eventually $p$ holds on all paths and $\A[p \WW q]$ means that $p$ holds on all paths until $q$ holds. Similar for operators $\EF$ and $\EE[p \WW q]$, which existentially quantify over paths. 
 We use $\FF$ and $\WW$ as our base temporal operators, and assume that formulae are written in negation normal form, in which negation only accurs next to atomic propositions, a formula that is not in negation normal form can be easily normalized. Note that $\AG p$ can be defined as $\A \lbrack p ~\WW~ \false \rbrack$, and $\EG p$ can be defined as $\EE \lbrack p ~\WW~ \false \rbrack$. 


\section{Dual Refinement}
\label{sec:refinement}
We now describe a method that, for a given NLA branch expression $b$ 
at some location $\loc$ of the program,
in a statement of the form $\ttif\ (\bloc)\ \ttthen\ s_1\ \ttelse\ s_2$,
will attempt to synthesize an equivalent Boolean LIA expression, using a combination of static and dynamic analysis.

\begin{figure}[t]
\begin{lstlisting}[language=customC, basicstyle=\ttfamily\footnotesize]
procedure $\dyRefine$($b$, $\synthposb$, $\synthnegb$):
  loop
    case |\TrimPos|: |\hspace{5pt}| $\exists \sigma.\ \sem{\synthposb}\sigma  \wedge \neg \sem{b}\sigma$ $\Rightarrow$ $\synthposb$ := $\synthposb \wedge \lnot formula(\sigma)$ |\ignore{\synthposb, \ \cexpath, false)}|
    case |\ExpandPos|: $\exists \sigma.\ \neg \sem{\synthposb}\sigma  \wedge \sem{b}\sigma$ $\Rightarrow$ $\synthposb$ := $\synthposb \vee formula(\sigma)$|\ignore{$\synthposb,\ \cexpath$, true)}| 
    case |\TrimNeg|:  |\hspace{-6pt}|  $\exists \sigma.\ \sem{\synthnegb}\sigma \wedge \sem{b}\sigma$ $\Rightarrow$ $\synthnegb$ := $\synthnegb \wedge \lnot formula(\sigma)$|\ignore{\dyGeneralize$\synthnegb,\ \cexpath$, false)}|
    case |\ExpandNeg|: $\exists \sigma.\ \neg \sem{\synthnegb}\sigma \wedge \neg \sem{b}\sigma$$\Rightarrow$ $\synthnegb$ := $\synthnegb \vee formula(\sigma)$|\ignore{$\synthnegb,\ \cexpath$, true)}|
	else $\Rightarrow$ return $(\synthposb,\synthnegb)$
\end{lstlisting}
\caption{\label{fig:dyrefine} Algorithm $\dyRefine$: Overall strategy synthesize an alternative to Boolean condition $b$ by refining a pair of conditions $\synthposb,\synthnegb$, so that  $\synthposb$ captures the conditions where $b$ holds and $\synthnegb$
captures the conditions where $\neg b$ holds.}
\end{figure}

\begin{figure}[t]
\includegraphics[width=5in]{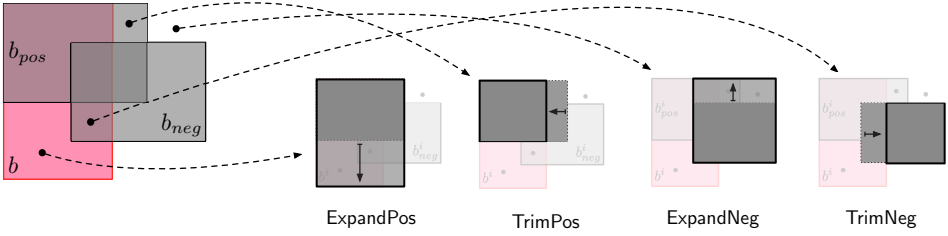}
\caption{Depictions of how candidate LIA conditions $\synthposb$ and $\synthnegb$
align with states where $b$ holds (in pink) and what actions are needed to remedy.}\label{fig:cases-actions}
\end{figure}

A key enabling insight is that, rather than synthesizing a single 
expression alternative to $b$, we can better explore both the positive
and negative sides of the state space (\ie~where $b$ holds and where it does not hold) by
synthesizing two expressions $\synthposb$ and $\synthnegb$, respectively.
By synthesizing $\synthposb$ and $\synthnegb$ together, we can 
track what we have learned about regions where $b$ holds and where $\neg b$ holds,
through conjunctive or disjunctive updates to $\synthposb$ or $\synthnegb$.
With this in mind, the refinement strategy that we perform is described in the algorithm in Fig.~\ref{fig:dyrefine}. 
From a candidate pair of conditions $\synthposb$ and $\synthnegb$, the algorithm iteratively checks to see whether one of four possible cases still hold. 
The cases are also represented pictorially in Fig.~\ref{fig:cases-actions}. 
\begin{itemize}
\item \TrimPos: In this case $\synthposb$ captures some states for which $b$ does not hold. For example, if $b \equiv x^2 > 4$ and $\synthposb \equiv x>1$ there is at least one state such as $x=2$, where  $\synthposb$ holds, but $b$ does not. In this case we need to reduce the size of $\synthposb$, as depicted in the first case of Fig.~\ref{fig:cases-actions}, so that it does not go beyond the pink $b$ area.
This will be accomplished by discovering some new Boolean formula subexpression (\S\ref{sec:dygeneralize} we introduce a method for this called $\dyGeneralize{}$) and conjunctively add it to $\synthposb$. For example, if $\dyGeneralize{}$ returns an expression equivalent to $x \neq 2$, then the conjunction is $\synthposb \equiv x \neq 2 \wedge x>1$, which can be simplified to $\synthposb \equiv x>2$.

\item \ExpandPos: In this case $\synthposb$ does not capture all of the states for which $b$ holds. For example, if $b \equiv x > 0$ and $\synthposb \equiv x > 5$, then there is a state such as $x=2$ where $b$ holds, but $\synthposb$ does not. In this case we need to refine by ``expanding'' $\synthposb$ to include the $x=2$ (and possibly other) states. This is depicted with the gray $\synthposb$ box expanding to cover more of the pink $b$ area.
In order to expand we again find a formula that represents $\sigma$ (and ideally other similar states) and add it disjunctively to $\synthposb$. 

\item \TrimNeg: In this case  $\synthnegb$ captures some states for which $\neg b$ does not hold. Here we need to shrink the gray  $\synthnegb$  box to reduce its overlap with the pink $b$ box.

\item \ExpandNeg: In this case $\synthnegb$ does not capture all of the states for which $\neg b$ holds. As depicted in the diagram, we need to enlarge the gray $\synthnegb$ box so that it covers states that are outside the pink box.
\end{itemize}

One simple optimization of this procedure is, while trimming the positive, use the same $\sigma$ to expand the negative (and similar for the expanding the positive while trimming the negative).

The above procedure is \emph{sound} in the sense that:
$   \forall\sigma. \sem{\synthposb}\sigma = \sem{b}\sigma \wedge \sem{\synthnegb} = \sem{\neg b}\sigma $. It is easy to see that procedure $\dyRefine$ will only return such a $(\synthposb,\synthnegb)$.
We discuss incompleteness below.

\dyRefine{} takes as argument initial guesses for $\synthposb,\synthnegb$.  Any initial guess is sound, but we use dynamic analysis to learn conditions that are close to $b$ and $\neg b$, respectively. To this end we use the instrumentation in \S\ref{sec:prelim}: for a given $\bloc$ in the program, we apply the instrumentation $\Psnap{}[\loc]$, execute the program under random inputs, collect
the sets of states $\snap{\loc,pos}$ and $\snap{\loc,neg}$, and then employ a learning routine $\textsf{learn}$ on each set to obtain conditions $\synthposb,\synthnegb$.

While the above algorithm is simple, critical to the practical success of the algorithm are the strategies needed for many subcomponents of the algorithm.
First, the semantics of condition $b$ depends on its context in the program, what invariants must hold of other variables, etc. To that end, in \S\ref{sec:validate} we discuss how counterexamples (\ie~$\sigma$) can be obtained that account for the program context, through a static transformation that reduces the problem to reachability and returns a counterexample from which $\sigma$ can be derived.
Second, individual counterexamples do not allow the refinement algorithm to make much progress on each iteration. Therefore in \S\ref{sec:dygeneralize} we describe a method of generalizing counterexamples using dynamic analysis.
There are also other details that arise at the implementation level, such as how to exploit the \emph{convex hull} when forming the disjunction in the refinement algorithm. We discuss these issues in the next sections.

\emph{Classes of input NLAs and synthesized LIA expressions.}
On the input side, our implementation does not enforce any specific class of non-linear expressions, apart from what is expressible in the C input programming language, and what the dynamic and static tools can support. As discussed in~\cite{tosem2013}, DIG uses dynamic analysis to analyze programs with polynomial expressions and infer nonlinear equalities and octagonal inequalities. Meanwhile, Ultimate~\cite{ultimate} parses input C programs into Boogie programs, and has some support for reachability of polynomials, a fact that we exploit in our static validation in \S\ref{sec:validate}.

On the output side, as one can see, the $\dyRefine$ algorithm synthesizes Boolean combinations of LIA equalities/inequalities or, formally:
$$
blia \;\;::=\;\;
blia \wedge blia' \mid
blia \vee blia' \mid
\neg blia \mid
f_{LIA}(x_1,\ldots,x_n) \varolessthan 0
$$
where $\varolessthan$ can be either $=$ or $<$, variables $x_1,\ldots,x_n \in \vars$ and
$f_{LIA}$ is a linear integer arithmetic expression over those variables.

Soundness of $\dyRefine$ depends on static validation, so we return to it at the end of the next section.
Another natural question is termination of refinement. This partially depends on validation (\S\ref{sec:validate}) as well as counterexample generalization
(\S\ref{sec:dygeneralize}), so we return to a discussion of termination in \S\ref{sec:termination}.

\section{Static Validation through Reachability}
\label{sec:validate}

Due to the unsoundness of dynamic analysis, the candidate linear conditions
$\synthposb$ and $\synthnegb$ of the non-linear conditions $b$ and $\neg b$, resp., must be statically validated against all executions.
Interestingly, $b$ and $\synthposb$ as well as $\neg b$ and $\synthnegb$ 
need not be equivalent in general where all models of their variables are considered.
It is only necessary for them to be equivalent in the reachable program states in which $b$ and $\neg b$ are evaluated, \ie~in their program contexts. For example, the condition \ttt{z*z - 12*y - 6*z + 12 + c <= k} in the running examples is not equivalent to its linear version \ttt{c <= k} in general (\eg~when $\ttt{y} = 0, \ttt{z} = 0$) but they are equivalent in the program context where the loop invariant $\ttt{z*z - 12*y - 6*z + 12} = 0$ holds. 

In this section we describe how to validate these candidate conditions in their program context, via a program transformation that reduces the problem to reachability.
When the candidates are not valid for all states, our transformation will also 
determine what kind of refinement is needed as a next step for $\synthposb$ and $\synthnegb$.

We define a \emph{replacement mapping}
$m : \locs \rightarrow \bstmt \times \bstmt$ 
to map a program location $\loc$ of a condition $\bloc$
to a pair of conditions $(\synthposbi, \synthnegbi)$ that, respectively, are candidate replacements for the conditions $\bloc$ and $\neg \bloc$.
We now transform the original program $P$ according to the mapping $m$ into a program $\Pcheck{}[m]$ by introducing some error locations in it for the following reason: if the program $\Pcheck{}[m]$ is safe (\ie\ no errors in $\Pcheck{}[m]$ are reachable) then all approximating positive and negative conditions $(\synthposbi, \synthnegbi)$ in $m$ are exact. The program transformation is defined as follows:

	$$
	\Pcheck{}[m]\ \  \defeq\ \  
	\left\{\begin{array}{l}
		\forall \loc \mapsto (\synthposbi, \synthnegbi) \in m.\\
		\ttif\ \bloc\ \ttthen\ s\ \ttelse\ s'
		\leadsto\\
		\;\;\;\;\;\;\left\{\begin{array}{lllll}
			\ttif & \bloc &\ttand\ &\neg \synthposbi & \ttthen \texttt{ goto }\errorPosTooS^\loc\\
			\ttelse\ttif\ & \bloc &\ttand\ &\synthnegbi & \ttthen \texttt{ goto }\errorNegTooB^\loc\\
			\ttelse\ttif\ & \neg \bloc &\ttand\ &\synthposbi & \ttthen \texttt{ goto }\errorPosTooB^\loc\\
			\ttelse\ttif\ & \neg \bloc  &\ttand\ &\neg \synthnegbi & \ttthen \texttt{ goto }\errorNegTooS^\loc\\
			\ttelse\ttif\ & \bloc\ &&&  \ttthen\ s\ \ttelse\ s'
		\end{array} \right\}
	\end{array}
	\right\}
	$$

Intuitively, the above transformation replaces each occurrence of an 
$\ttif\ \bloc\ \ttthen\ s\ \ttelse\ s'$ with 5-way branching. 
The final branch involves the normal control-flow of the program, allowing executions of the original program to also exist in the transformed program. 
However, the first four branches test all of the ways in which $\bloc$ can be inconsistent with $\synthposbi$ and $\synthnegbi$.
If it is possible for an execution of the original program $P$ to reach location $\loc$
in a state where such an inconsistency holds,
then there will be an execution of $\Pcheck{}[m]$ that can reach the corresponding error label for that form of inconsistency.

\begin{example}
Fig.~\ref{fig:instr-error} shows the original running example discussed in \S\ref{sec:overview} again on the left. To the right is the resulting transformed $\Pcheck{}[m]$ when using $m=\{\loc_{\ref{ln:expoly}} \mapsto (\texttt{0>=c-k,0=c-k-1})\}$, which maps the polynomial on line~\ref{ln:expoly} to the current candidate $\synthposb,\synthnegb$.
There will be an execution of this transformed program that enters the branch on Line~\ref{ln:instr-ps} if ever there is a reachable program state where
the polynomial inequality holds, but $\synthposb$ does not. In such a circumstance a reachability 
analysis will emit a counterexample that reaches $\errorPosTooS$, reflecting that $\synthposb$ needs to be expanded. The next 3 branches are similar. 
If an execution does not fall into one of the first four branches, this does not (yet) mean that $\synthposb,\synthnegb$ are valid: it could, \eg~be that a state at a later loop iteration witnesses a shortcoming of $\synthposb$ or $\synthnegb$. Thus, the fifth branch allows executions to fall through the check and continue to later program states. 
If no execution can ever reach any of the error labels, then $\synthposb$ and $\synthnegb$ must be accurate, \ie~the soundness condition discussed further below.

Let us now look at a counterexample to the validity of $\Pcheck{}[m]$ and see that it indicates how $\synthposb$ or $\synthnegb$ needs to be amended. Recall that the counterexample is in the form of a feasible sequence of program statements that lead to the error location, such as the following (for the instrumented program on the right in Fig.~\ref{fig:instr-error}):
\begin{equation}\label{eqn:cex1}
cex_1 \equiv \left\{
 \begin{array}{rl} 
   \loc~\ref{ln:instr-init}:  & \ttt{int y=1, z=6, c=0, p=2;} \\
   \loc~\ref{ln:instr-init2}: & \ttt{int k=*;}\\
   \loc~\ref{ln:instr-while}: &  \ttt{assume(true);}\\
   \loc~\ref{ln:instr-pb}:    &  \ttt{assume(!(z*z - 12y - 6z + 12 + c <= k));}\\
   \loc~\ref{ln:instr-pb}:    &  \ttt{assume(!(0 >= c-k));}\\
   \loc~\ref{ln:instr-ns}:    & \ttt{assume(!(1 = c-k);}\\
     \loc~\ref{ln:instr-nsb}: &  \errorNegTooS\\
\end{array}
\right.
\end{equation}
This is a feasible program path. Consider, for example, the case where $\ttt{k}=-2, c=0$ initially and, in that case
both $b$ and $\synthnegb$ ($c-k-1=0$) are $\ttt{false}$. Therefore branch condition (\texttt{$\neg b$ \&\& $\lnot\synthnegb$}) holds so error location $\errorNegTooS$ can be reached and the above program path $cex_1$ is output by a reachability verifier, with the error location indicating that $\synthnegb$ must be expanded.
\end{example}

\begin{figure}[t]
\begin{minipage}[t]{0.36\textwidth}
\begin{lstlisting}[language=customC,style=pnnstyle,basicstyle=\ttfamily\footnotesize]
int y=1, z=6, c=0, p=2;   
int k=*;
while (true):








  if (z*z-12y-6z+12+c>k):|\label{ln:expoly}|
    break;
  else:
    y = y + z;
    z = z + 6;
    c = c + 1;
    p = 1;
p = 0;
return 0;
\end{lstlisting}
\end{minipage}
\hspace{-15pt}
\begin{minipage}[t]{0.64\textwidth}
\begin{lstlisting}[language=customC,style=pnnstyle,numbers=none,basicstyle=\ttfamily\footnotesize]
int y=1, z=6, c=0, p=2; |\label{ln:instr-init}|
int k=*; |\label{ln:instr-init2}|
while(true):|\label{ln:instr-while}|
  if (z*z-12y-6z+12+c<=k&&$\lnot$($\color{codegreen}{\synthposb}: $0$\geq$c-k)):|\label{ln:instr-ps}|
    $\errorPosTooS$ // ${\color{codegreen}\synthposb}$ too small|\label{ln:instr-psb}|
  elsif (z*z-12y-6z+12+c<=k&&($\color{codegreen}{\synthnegb}: $0$=$c-k-1))):|\label{ln:instr-nb}|
    $\errorNegTooB$ // ${\color{codegreen}\synthnegb}$ too big|\label{ln:instr-nbb}|
  elsif ($\neg$(z*z-12y-6z+12+c<=k)&&($\color{codegreen}{\synthposb}: $0$\geq$c-k))):|\label{ln:instr-pb}|
    $\errorPosTooB$ // ${\color{codegreen}\synthposb}$ too big|\label{ln:instr-pbb}|
  elsif ($\neg$(z*z-12y-6z+12+c<=k)&&$\lnot$($\color{codegreen}{\synthnegb}: $0$=$c-k-1))):|\label{ln:instr-ns}|
    $\errorNegTooS$ // ${\color{codegreen}\synthnegb}$ too small|\label{ln:instr-nsb}|
  elsif (z*z-12y-6z+12+c>k):|\label{ln:instr-if}|
    break|\label{ln:instr-br}|
  else:|\label{ln:instr-else}|
    y = y + z;|\label{ln:instr-b1}|
    z = z + 6;|\label{ln:instr-b2}|
    c = c + 1;|\label{ln:instr-b3}|
    p = 1;|\label{ln:instr-b4}|
p = 0;|\label{ln:instr-r1}|
return 0;|\label{ln:instr-r2}|
\end{lstlisting}
\end{minipage}
\caption{Demonstration of the $\Pcheck{}[m]$ transformation on the program from \S\ref{sec:overview}.}\label{fig:instr-error}
\end{figure}

If $\Pcheck{}[m]$ is safe, then we have the following guarantee. (Proof sketch in Apx.~\ref{apx:soundness}).

\begin{lemma}[Transformation Correctness]\label{lemma:sound}
	If all errors in $\Pcheck{}[m]$ are unreachable then 
	\[
	\forall \loc \mapsto (\synthposbi, \synthnegbi) \in m.\,\forall \tstate{} \in \preds{\bloc}.\,
	\sem{\synthposbi}\tstate{} = \sem{\bloc}\tstate{} \wedge \sem{\synthnegbi}\tstate{} = \sem{\neg \bloc}\tstate{} 
	\]
\end{lemma}

\paragraph{Effectiveness of static validation.}
We began from the premise that static validation techniques do not cope well with polynomials, yet we now find ourselves using them for exactly that purpose in $\Pcheck{}[m]$. There are a few reasons why this is not a sleight of hand. 
First, we do not ask a static tool to discover a linear replacement for a polynomial but instead validate one that was obtained through dynamic learning.
Second, we do not need to reason perfectly about the polynomials that occur inside the Boolean $\bloc$'s in $\Pcheck{}[m]$ above but instead only need to reason about the Boolean properties of the polynomials.
Finally, as mentioned above, we also do not need to reason perfectly about the polynomials on all inputs but instead only in the program context (reachable states) where they occur.
In \S\ref{sec:eval}, we will see that tools such as Ultimate~\cite{ultimate} are indeed able to validate  $\Pcheck{}[m]$, despite these instances of polynomial Boolean expressions.

\section{Dynamic Generalization of Counterexamples}
\label{sec:dygeneralize}

A counterexample to the static validation ($\Pcheck{}[m]$ in the previous section), including concrete witness values for the variables (a model), demonstrates the 
insufficiency of  the current candidate linear approximating conditions $\synthposb$ and $\synthnegb$. As seen above, this counterexample from $\Pcheck{}[m]$ corresponds to a path in the original program $P$ leading to a place where
the truth value of some $\bloc$ is inconsistent with $\synthposbi$ or $\synthnegbi$.
Already the counterexample model could be used to assist in refining $\synthposb$ and $\synthnegb$. 
Unfortunately, a single such model would only refine $\synthposbi$ or $\synthnegbi$ by a single data point, which would not lead to a tractable overall algorithm. In infinite domains, the refinement process may diverge if we refine with only concrete error snapshots.

In this section we introduce $\dyGeneralize$ (shown in Fig.~\ref{fig:dygeneralize} and discussed below), a technique to employ dynamic learning to discover a broader \emph{condition} denoted $b_{cex}$ that is learned from many counterexample models, and allows our overall $\dyRefine$  to take more significant steps toward completion. $\dyGeneralize$ is called with the current $\synthposb/\synthnegb$ needed amendment, the current counterexample path $cex$ and the \lstinline{Expand/Trim} direction of amendment. There are then three steps:

{\bf Step 1.}
$\dyGeneralize$ first generates many concrete snapshots (\eg~a parameter value of 1,000) at the error location from the input counterexample. This can be seen on Line~\ref{ln:getmodels-formula} in Fig.~\ref{fig:dygeneralize}.  
\begin{wrapfigure}{r}{0.25\textwidth}
  \vspace{-0.2in}
  \small
  \begin{tabular}{lllllll}
    \ttt{y} & \ttt{z} & \ttt{c} &  \ttt{k} & \ttt{p}\\
    \hline
    1 & 6 & 0 &  -2 & 2\\
    1 & 6 & 0 &  -3 & 2\\
    1 & 6 & 0 &  -4 & 2\\
    1 & 6 & 0 &  -5 & 2\\
    1 & 6 & 0 &  -6 & 2\\
    \multicolumn{5}{c}{...}
  \end{tabular}
  \vspace{-0.15in}
\end{wrapfigure}
Our procedure $\dyGeneralize$ transforms a counterexample path
into an SSA logical formula using standard techniques, denoted \lstinline{path2formula}($cex$). From such formula, we can harvest distinct error snapshots by iteratively invoking an SMT solver to get a model at the error location and then adding constraints to to prevent the solver generating the same model in the future iterations. 
For example, the table in the right shows some error snapshots extracted from counterexample $cex_1$ (Eqn.~\ref{eqn:cex1}).

{\bf Step 2.}
Next, $\dyGeneralize$ employs dynamic analysis to learn an error condition $b_{cex}$ from the collected error snapshots (Line~\ref{ln:learn} in Fig.~\ref{fig:dygeneralize}). 
From the data points in the above right table, dynamic analysis can learn 
the condition $b_{cex} \equiv 0 \geq p + k \wedge  0 = p - 2 \wedge 0 = c$.

{\bf Step 3.}
Depending on the refinement action determined from the counterexample, a simpler and more useful condition can be extracted from the error condition. In particular, if the counterexample indicates that an expansion refinement is needed, the current approximating condition $b_{cur}$ should be expanded with only new conditions which cover program states not covered by $b_{cur}$.  Because such new conditions in $b_{cex}$ contradict $b_{cur}$, they can be identified from the unsatisfiable core of $b_{cur} \wedge b_{cex}$, as seen on Line~\ref{ln:usc}. The exact way the unsatisfiable core is utilized is described below.

\begin{figure}[t]
\begin{lstlisting}[language=customC, basicstyle=\ttfamily\footnotesize]
procedure $\dyGeneralize$($b_{cur}$, $\cexpath$, direction):
	$S$ := getModels(path2formula($\cexpath$), iters=1000);|\label{ln:getmodels-formula}| 
	$b_{cex}$ := $\texttt{learn}$($S$);|\label{ln:learn}| 
	if (direction == Expand):
		match  UnsatCorePair$(b_{cur}, b_{cex})$ with|\label{ln:usc}| 
		$\mid$ Some($b_{usc}$) $\Rightarrow$ return $b_{usc}$
		$\mid$ None $\Rightarrow$ return $b_{cex}$
	else:
		return $b_{cex}$
\end{lstlisting}
\caption{\label{fig:dygeneralize} Algorithm $\dyGeneralize$: Generalizing a single counterexample $cex$ beyond a single model, to a formula that captures many states that could reach the same counterexample location.}
\end{figure}

\begin{definition}[Unsatisfiable core pairs] Given $\varphi$, $\varphi_1$ and $\varphi_2$ in CNF as a set of clauses, and assuming a $\ttt{UnsatCore}(\varphi)$ that returns an unsatisfiable subset of the set of clauses in $\varphi$, if one exists, then $\ttt{UnsatCorePair}$, given $\varphi_1$ and $\varphi_2$, is defined as the set of clauses in $\ttt{UnsatCore}(\varphi_1 \cup \varphi_2) \cap \varphi_2.$
\end{definition}

\begin{figure}[h]
\begin{lstlisting}[language=customC, basicstyle=\ttfamily\footnotesize]
procedure $\convexhullOR$($b$, $b'$):
   match convexHull($b \vee b'$) with
   $\mid$ Some($b''$)  $\rightarrow$ return $b''$
   $\mid$ None  $\rightarrow$ return $b \vee b'$

procedure $\dyRefine'$($b$, $\synthposb$, $\synthnegb$):
loop
  case |\TrimPos|: |\hspace{5pt}| $\exists cex.\ {\synthposb}  \wedge \neg {b}$ $\rightarrow$ $\synthposb$ := $\synthposb \wedge \dyGeneralize$($\synthposb$, $cex$, Trim)
  case |\ExpandPos|: $\exists cex.\ \neg {\synthposb}  \wedge {b}$ $\rightarrow$ $\synthposb$ := $\synthposb \convexhullOR \dyGeneralize$($\synthposb$, $cex$, Expand)
  case |\TrimNeg|:  |\hspace{-6pt}|  $\exists cex.\ {\synthnegb} \wedge {b}$ $\rightarrow$ $\synthnegb$ := $\synthnegb \wedge \lnot \dyGeneralize$($\synthnegb$, $cex$, Trim)
  case |\ExpandNeg|: $\exists cex.\ \neg {\synthnegb} \wedge \neg {b}$ $\rightarrow$ $\synthnegb$ := $\synthnegb \convexhullOR \dyGeneralize$($\synthnegb$, $cex$, Expand)
  else $\Rightarrow$ return $(\synthposb,\synthnegb)$
\end{lstlisting}
\caption{\label{fig:dyrefine2} Algorithm $\dyRefine'$: A revised version of $\dyRefine$ from Fig.~\ref{fig:dyrefine} that now employs dynamic counterexample generalization, and uses the convex hull for disjunction.}
\end{figure}

{\bf Putting it all together.}
We now describe a revised $\dyRefine'$, shown in Fig.~\ref{fig:dyrefine2} that employs this $\dyGeneralize$. First, we use a counterexample path $cex$ (rather than a single model $\sigma$ in Fig.~\ref{fig:dyrefine}), which is passed to $\dyGeneralize$, along with the current condition to be modified, and the appropriate \lstinline{Expand/Trim} directive.
In the \lstinline{Expand} case, the condition returned by $\dyGeneralize$ is used to form a disjunction with the current condition $b_{cur}$. 
Although it could be used immediately, we can further approximate that disjunction with its convex hull,
denoted $\convexhullOR$ and defined in Fig.~\ref{fig:dyrefine2}, for a faster converging refinement. For instance, in the running example, consider the refinement when the counterexample determines that the current condition $\synthnegb \equiv  c -k = 1$ can be expanded with the condition $b_{cex} \equiv 0 \geq p + k \wedge  0 = p - 2 \wedge 0 = c$. The conditions $\synthnegb$ and $b_{cex}$ almost cover every possible program state in which the valuation of $\neg b \equiv c > k$ is \ttt{true}, \ie\ (i) when $k$ is initially non-negative and the loop executes some iterations before terminating by the condition $c = k - 1$ and (ii) when $k$ is initially negative (the state $c = 0, k = - 1$ is not covered in $b_{cex}$) and  the loop immediately terminates without any iteration. Despite that, their disjunction $\synthnegb \vee b_{cex} \equiv  c = k + 1 \vee 0 \geq p + k \wedge  0 = p - 2 \wedge 0 = c$ is still missing some program states from the desired condition $\neg b$ (\eg\ the state $c = 0, k = - 1$) which requires more refinement steps. It also contains the variable $p$ which is irrelevant to the result. Fortunately, the convex hull of this disjunction is exactly $c > k$ so that no further refinement steps are needed.


\section{Termination/Divergence of Refinement}
\label{sec:termination}

Recall that the output of our algorithm is 
a Boolean combination of linear integer arithmetic equalities/inequalities
($blia$ as defined in Sec.~\ref{sec:refinement}).
However, expressing arbitrary polynomial equalities as a $blia$ is not always feasible, much less whether our algorithm would always discover one.
Nonetheless, as seen in Sec.~\ref{sec:eval}, our algorithm frequently does discover such $blia$ conditions and we now discuss why, identify fragments where termination is guaranteed and identify sources of divergence.

\emph{Termination due to interval analysis.}
Perhaps surprisingly, we often do discover a $blia$ equivalent of the original 
NLA. The main reason is that we only need to capture the Boolean aspects of the 
polynomial, which amounts to knowing when the polynomial will be above or below a certain bound.
\begin{wrapfigure}[8]{r}[34pt]{3.2in}
\includegraphics[width=2.7in]{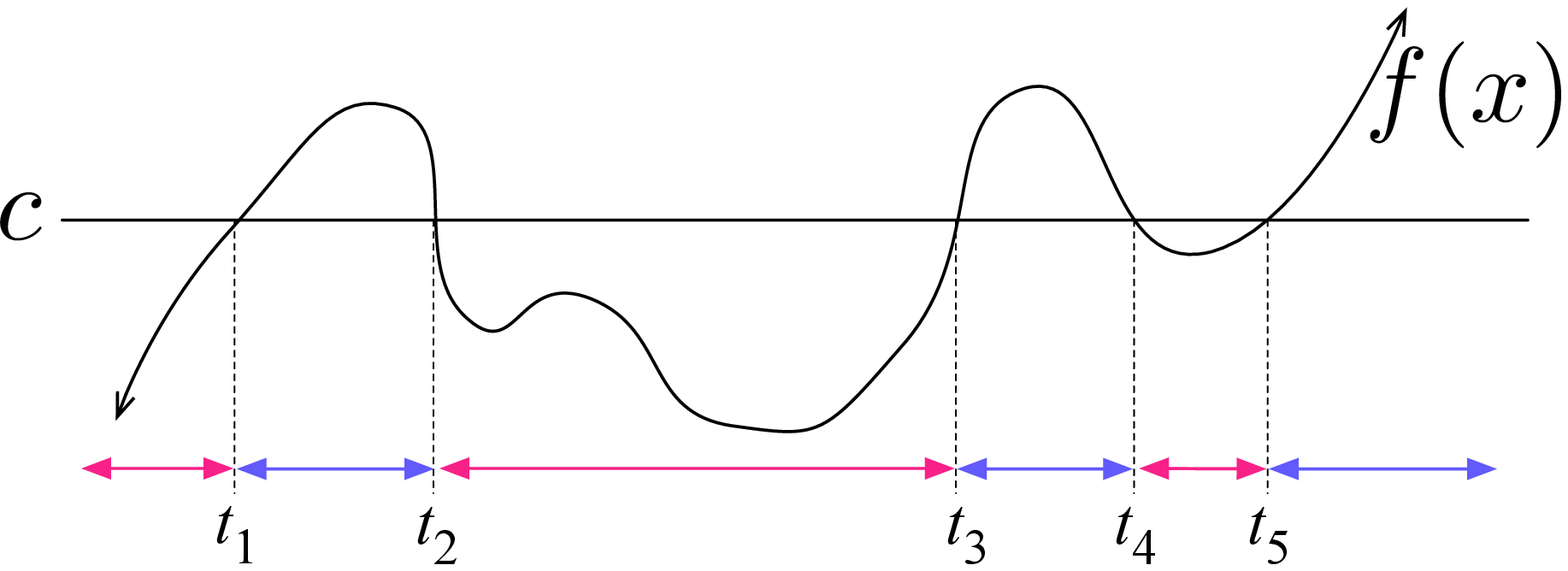}
\end{wrapfigure}
Consider the diagram to the right.
In this single-variable example of some polynomial $f(x)$ and Boolean property
$f(x) < c$, an equivalent $blia$ need only capture when $f(x)$ is above or below $c$. 
That is, the $blia$ simply needs to distinguish the blue values of $x$ 
(when $f(x) \geq c$) from the red values of $x$ (when $f(x) < c$), and this can be achieved with the expressions:\\
\[\begin{array}{lll}
b_{pos} \equiv (t_1 < x \wedge x < t_2) \vee (t_3 < x \wedge x < t_4) \vee (t_5 < x),\\
b_{neg} \equiv (x \leq t_1) \vee (t_2 \leq x \wedge x \leq t_3) \vee (t_4 \leq x \wedge x \leq t_5).
\end{array}\]
These $blia$ expressions are essentially interval constraints~\cite{intervalAbsDom}.
Note that there is a potential application here toward compiler optimization, if the calculation of such intervals is more efficient than calculating the polynomial, though we leave this to future work.
A simple example of the interval phenomenon is an inequality such as
$x^2 > 49$, where an equivalent $blia$ is a simple disjunction of intervals.
Beyond a single variable, the two variable inequality
$x^2 + y^2 < 4$ has a $blia$ alternative of
$(-2 < y \wedge y < 2) \wedge (-2 < x \wedge x < 2)$.

\emph{Termination in other special cases.}
In some program contexts, polynomial expressions may always evaluate to a constant amount.
For example, if an NLA such as $a^2-b^2+y+5>0$
occurs inside a loop and an invariant of the loop is
that $y = 0 \wedge a = b$, then the NLA will always be equivalent to $5>0$.
Slightly more generally, 
in some program contexts, polynomial expressions may always be directly equivalent to
an LIA equality/inequality. For example, in a loop where 
$a = - b$ is an invariant,
the polynomial inequality
$(a + b)^2 + x > 0$
is exactly equal to
$x > 0$.
An example is the program in Sec.~\ref{sec:overview}, where a portion of the polynomial is equivalent to $0$ due to loop invariants.

\emph{Divergence.}
Finally, there are cases where divergence is inevitable.
First, as noted, there may be polynomials that simply cannot be expressed as a $blia$.
Second, practically speaking, our dynamic learning in DIG may not be able to learn a sufficiently precise LIA expression for us to use as a building block.
Finally, in some cases the dynamic generalization may cause us to ``over-shoot,'' generalizing a counterexample to create a $b_{cex}$ that encompasses an important interval stopping point, \eg~we may generalize data points
$x=25$ and $x=50$ to $b_{cex}=x>20$, despite there being an important interval at $x=51$. In future work, we aim to detect those circumstances and attempt a binary search strategy to iteratively reduce the generalization ranges. 
Practically, to avoid divergence, our implementation currently forces termination after a fixed number of steps (18 iterations) and returns the partially synthesized $\synthposb$ and $\synthnegb$.


\section{Implementation}
\label{sec:impl}

We implemented dual rewriting in a new
tool called \Tool, written in a combination of Python and OCaml. \Tool\ uses DIG~\cite{tosem2013} for dynamic learning and Ultimate~\cite{ultimate} in reachability mode for static validation.
The algorithm takes an input program $P$ with NLA expressions and emits a mapping $m : \locs \rightarrow (e,e')$ mapping expression locations in $P$ to a pair of LIA conditions $b,b'$ that can replace those NLAs.
The following is a diagram of \Tool{}:\\
\begin{center}
 \includegraphics[width=0.80\textwidth]{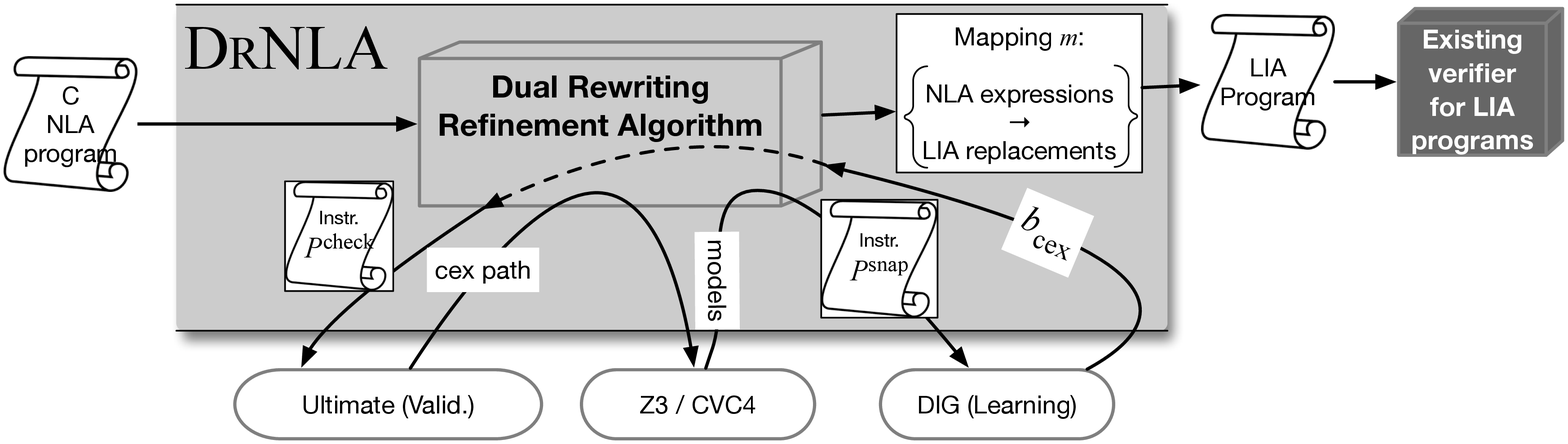} 
\end{center}
\Tool's main driver is implemented in Python.
During the refinement loop, \Tool~calls DIG and Ultimate iteratively, and \Tool{} uses OCaml/CIL to construct the $\Psnap{}$ instrumentation for DIG and $\Pcheck{}$ instrumentation for Ultimate.
Ultimate returns counterexample paths, which we parse and convert to SSA, so that we can query Z3 to generate many models. We then use DIG to learn from those models a new expression $b_{exp}$ which we parse, add to $\synthposb/\synthnegb$, and reincorporate into the current map $m$ to be validated again.

\dig{}'s learning procedure is used in two circumstances: first to infer initial guess for  $\synthposb$/$\synthnegb$ and second during generalization of counterexamples.
The learning procedure actually returns a \emph{list} of candidate invariants and not all of them are useful. In fact, we have two lists:
one for $\synthposb$ and one for $\synthnegb$.  We start by removing identical invariants which are useless in discriminating $\synthposb$ from $\synthnegb$. We then 
prune away irrelevant invariants using an UNSATcore procedure to return the minimal sets of candidate invarants. Our dynamic analysis instrumentation also has to take care that the programs do not run forever, so we also instrument loop bounds.


\section{Evaluation}\label{sec:eval}

\newcommand{\unsound}[1]{\Lightning\!\! #1}
\newcommand*{\hl}{\cellcolor{black!10}}

\newcommand*\benchtablerowstretch{0.9}
\newcommand*\benchtabletabcolsep{0.2em}

\newcommand*{\hlg}{\cellcolor{green!20}}
\newcommand*{\rhl}{\rowcolor{black!10}}

\newcommand*\OK{\ding{52}\xspace}
\newcommand*\rTO{\textbf{TO}\xspace}
\newcommand*\rOOM{\textbf{MO}\xspace}
\newcommand*\NOK{\ding{55}\xspace}
\newcommand*\FAIL{\Radioactivity\xspace}
\newcommand*\rCRASH{\Radioactivity\xspace}
\renewcommand*\rUNK{{\textbf{?}\xspace}}
\renewcommand\rFALSE{\NOK}

\newcommand\rExact{\OK}
\newcommand\rAppx{$\approx$}

We evaluated whether \Tool\ (i) was able to synthesize equivalent LIA Boolean conditions from NLA conditions and (ii) whether these synthesized LIA expressions, when replaced in the program, enable existing tools to verify properties of previously unsolvable problems.  In this paper we apply our dual rewriting idea on verifying branching time (CTL) programs, both because of the popularity and practicality of CTL properties and the state-of-the-art CTL tools, e.g., T2 and \FuncTion{}, have strong support for LIA but not NLA expressions and properties. We also demonstrate a potential use of \tool{} in program slicing using the Frama-C analyzer.

All experiments described below were run on an Amazon AWS cloud instance with 4 virtual 2.4GHz CPUs, 16GB of memory, and Ubuntu Linux.  \tool{} and all external tools (e.g., Ultimate, T2, \FuncTion{}) use a 900s timeout.
We run T2 and \FuncTion{} with default parameters (though we did try to use various parameters such as abstract domains in T2 but did not find ones that improve the performance of the tools).  In the supplementary materials, we have provided artifacts containing scripts to invoke the tools and collect results for reproducibility purposes.

\subsection{Benchmarks}\label{sec:benchmarks}
As our chosen application is on temporal verification, we sought for benchmark suites consisting of CTL properties for programs with NLA expressions. To our knowledge, there are no existing CTL+NLA benchmarks and we thus create such benchmarks using two complementary strategies:
extending NLA benchmarks to include CTL properties and  extending CTL benchmarks to include NLA expressions.

(1) {\benchsvcomp.} We first adapted the NLA benchmarks from the Dynamite termination work~\cite{le2020dynamite}, which came from the \texttt{nla-digbench} category if SV-COMP~\cite{nladigbench}.
These 28 programs contain NLA loop invariants and implement mathematical functions such as \texttt{intdiv}, \texttt{gcd}, \texttt{lcm}, and \texttt{powersum}.
We instrument these programs to include CTL properties. For each program with $n$ NLA loop invariants, we create $n$ variants of the program, where each has an NLA loop invariant and a new variable $p$ that is updated in existing loops in such a way that the variable is either 0 or 1 when the program exits.
Thus these programs have the CTL property $\EF(p=0) \land \EF(p=1)$.
Next, we create another $n$ variants of the program with the same NLA loop invariants but the variable $p$ is updated in such a way that the CTL property does not hold.
Table~\ref{tab:res1} shows the 56 benchmarks: the CTL property is valid for programs with suffix-T and invalid for those with suffix-F.

(2) {\benchpldi.} We also created 21 benchmarks with various types of CTL properties from those found in~\cite{CookK2013}. For each program, we insert a simple terminating loop that contains NLA expressions.
The goal of this instrumentation is to force the CTL tool to reason about the behaviors of these NLA expressions to determine the termination of the loop.
In addition, we select 5 programs randomly
from these benchmarks and insert several NLA loops (taken from \benchsvcomp) to increase the difficulty and also diversity of NLA expressions.
The second stanza of
Table~\ref{tab:res1} shows these 26 programs: the upper part lists the 21 original programs and the lower part lists the 5 additional programs.

(3) {\benchcustom.}
We also create a set of 10 more challenging handcrafted benchmarks to evaluate \tool{}, shown in the third stanza of Table~\ref{tab:res1}.
These are created with combinations of NLA expressions, which result in higher degrees and are used as conditional or loop guards.  The LIA equivalent expressions of these NLA expressions are  more complex and involve disjunctions of linear constraints.

In total we have 92 CTL benchmarks with NLA and CTL properties (56 from \benchsvcomp{},  26 from \benchpldi{}, and 10 from \benchcustom). Our modifications to the original benchmarks are relatively simple (\eg~simple CTL property used in \benchsvcomp), yet sufficiently strong to illustrate the limitations of existing CTL tools in dealing with NLA programs as shown Sec.~\ref{sec:results}. These benchmarks are available in the supplementary materials.

\subsection{Synthesizing LIA Replacements for NLA Expressions}\label{sec:results}

\begin{table}
  \small
  \caption{\tool{}'s rewrite results for \benchsvcomp{}, \benchpldi, and \benchcustom.}\label{tab:res1}  
  \begin{tabular}{ll|c|r|r||l|c|r|r}
    \toprule
\multirow{26}{0.1in}{\rotatebox[origin=c]{90}{{\bf Benchmarks from} \benchsvcomp}}&
    \textbf{Benchmark} & \textbf{Res} & \textbf{T(s)} & \textbf{It.} & \textbf{Benchmark} & \textbf{Res} & \textbf{T(s)} & \textbf{It.} \\    
    \midrule
&\ttt{bresenham1-F.c} & \rExact  & 210.1  & 2   & \ttt{fermat1-F.c} & \rAppx   & \rTO      & 0   \\
&\ttt{bresenham1-T.c} & \rExact  & 204.2  & 2   & \ttt{fermat1-T.c} & \rAppx   & \rTO      & 0   \\
&\ttt{cohencu2-F.c} & \rAppx   & 517.5    & 1   & \ttt{geo1-F.c}  & \rExact  & 131.0       & 1   \\
&\ttt{cohencu2-T.c} & \rAppx   & 473.7    & 1   & \ttt{geo1-T.c}  & \rExact  & 130.6       & 1   \\
&\ttt{cohencu3-F.c} & \rAppx   & 621.8    & 1   & \ttt{geo2-F.c}  & \rExact  & 122.5       & 1   \\
&\ttt{cohencu3-T.c} & \rAppx   & 621.4    & 1   & \ttt{geo2-T.c}  & \rExact  & 134.7       & 1   \\
&\ttt{cohencu4-F.c} & \rAppx   & 780.3    & 2   & \ttt{geo3-F.c}  & \rExact  & 171.1       & 1   \\
&\ttt{cohencu4-T.c} & \rAppx   & 773.9    & 2   & \ttt{geo3-T.c}  & \rExact  & 163.5       & 1   \\ 
&\ttt{cohencu5-F.c} & \rExact  & 179.4    & 2   & \ttt{hard-F.c}  & \rUNK    & 7.9         & 0   \\ 
&\ttt{cohencu5-T.c} & \rExact  & 177.4    & 2   & \ttt{hard-T.c}  & \rUNK    & 7.4         & 0   \\ 
&\ttt{cohencu7-F.c} & \rAppx   & 194.9    & 2   & \ttt{hard2-F.c} & \rUNK    & 0.9         & 0   \\
&\ttt{cohencu7-T.c} & \rAppx   & 667.5    & 2   & \ttt{hard2-T.c} & \rUNK    & 0.8         & 0   \\ 
&\ttt{dijkstra2-F.c} & \rAppx   & 687.5   & 2   & \ttt{prod4br-F.c} & \rAppx   & 239.0     & 5   \\ 
&\ttt{dijkstra2-T.c} & \rAppx   & 686.0   & 2   & \ttt{prod4br-T.c} & \rAppx   & 159.3     & 4   \\ 
&\ttt{dijkstra3-F.c} & \rAppx   & 628.6   & 1   & \ttt{prodbin-F.c} & \rAppx   & 654.5     & 2   \\
&\ttt{dijkstra3-T.c} & \rAppx   & 627.5   & 1   & \ttt{prodbin-T.c} & \rAppx   & 619.6     & 1   \\
&\ttt{dijkstra4-F.c} & \rAppx   & 629.7   & 1   & \ttt{ps2-F.c}   & \rExact  & 29.0        & 1   \\
&\ttt{dijkstra4-T.c} & \rAppx   & 629.1   & 1   & \ttt{ps2-T.c}   & \rExact  & 53.4        & 2   \\
&\ttt{dijkstra5-F.c} & \rAppx   & 630.8   & 1   & \ttt{ps3-F.c}   & \rExact  & 62.8        & 2   \\
&\ttt{dijkstra5-T.c} & \rAppx    & 613.9  & 1   & \ttt{ps3-T.c}   & \rExact  & 44.9        & 1   \\
&\ttt{divbin1-F.c} & \rUNK    & 1.2       & 0   & \ttt{ps4-F.c}   & \rExact  & 59.8        & 2   \\
&\ttt{divbin1-T.c} & \rUNK    & 1.3       & 0   & \ttt{ps4-T.c}   & \rExact  & 58.1        & 2   \\
&\ttt{egcd-F.c}  & \rAppx    & \rTO       & 0   & \ttt{ps5-F.c}   & \rExact  & 30.4        & 1   \\
&\ttt{egcd-T.c}  & \rAppx   & \rTO        & 0   & \ttt{ps5-T.c}   & \rExact  & 41.3        & 1   \\
&\ttt{egcd2-F.c} & \rAppx   & 696.0       & 1   & \ttt{ps6-F.c}   & \rExact  & 60.3        & 2   \\
&\ttt{egcd2-T.c} & \rAppx   & 681.8       & 1   & \ttt{ps6-T.c}   & \rExact  & 68.6        & 2   \\
&\ttt{egcd3-F.c} & \rAppx    & \rTO       & 8   & \ttt{sqrt1-F.c} & \rExact  & 117.0       & 2   \\
&\ttt{egcd3-T.c} & \rAppx   & \rTO        & 8   & \ttt{sqrt1-T.c} & \rExact  & 116.3       & 2   \\
    \bottomrule
\multirow{14}{0.1in}{\rotatebox[origin=c]{90}{{\bf Benchmarks from} \benchpldi}}&
\ttt{afagp-F.c} & \rExact  & 103.9    & 2  &   \ttt{afp-F.c}   & \rExact  & 163.8    & 2  \\
&\ttt{afagp-T.c} & \rExact  & 74.8     & 1  &   \ttt{afp-T.c}   & \rExact  & 166.8    & 2  \\
&\ttt{afefp-T.c} & \rExact  & 179.8    & 2  &   \ttt{agafp-F.c} & \rExact  & 65.7     & 1  \\
&\ttt{afegp-F.c} & \rExact  & 239.7    & 2  &   \ttt{agafp-T.c} & \rExact  & 36.3     & 1  \\
&\ttt{afegp-T.c} & \rExact  & 177.8    & 2  &   \ttt{efafp-T.c} & \rExact  & 231.1    & 2  \\
&\ttt{neg-afagp-F.c} & \rExact  & 289.1    & 2  & \ttt{neg-afp-F.c} & \rExact  & 164.4      & 2  \\
&\ttt{neg-afagp-T.c} & \rExact  & 287.2    & 2  & \ttt{neg-afp-T.c} & \rExact  & 160.9      & 2  \\
&\ttt{neg-afefp-F.c} & \rExact  & 225.2    & 2  & \ttt{neg-efafp-F.c} & \rExact  & 243.0    & 2  \\
&\ttt{neg-afegp-F.c} & \rExact  & 226.0    & 2  & \ttt{neg-egafp-F.c} & \rExact  & 282.2    & 2  \\
&\ttt{neg-afegp-T.c} & \rExact  & 217.4    & 2  & \ttt{neg-egafp-T.c} & \rExact  & 313.3    & 2  \\
&\ttt{neg-egimpafp-T.c} & \rExact  & 251.4 & 2 & &&& \\
&\ttt{afagp-T.c} & \rExact  & 160.3    & 1  &  \ttt{neg-afefp-F.c} & \rExact  & 233.2& 1  \\
&\ttt{afefp-T.c} & \rExact  & 312.1    & 2  &  \ttt{neg-afp-F.c} & \rExact  & 68.2   & 2  \\
&\ttt{agafp-T.c} & \rExact  & 24.4     & 1  &  &&&\\
    \bottomrule
\multirow{6}{0.1in}{\rotatebox[origin=c]{90}{\benchcustom}}&
\ttt{if-cubic-F.c} & \rExact  & 51.5     & 3  & \ttt{square-loop-F.c} & \rExact  & 383.6    & 12 \\ 
&\ttt{if-cubic-T.c} & \rExact  & 51.3     & 3  & \ttt{square-loop-T.c} & \rExact  & 233.3    & 7 \\  
&\ttt{if-F.c}    & \rExact  & 78.8     & 6     & \ttt{while-cubic-F.c} & \rExact  & 87.0     & 7  \\ 
&\ttt{if-T.c}    & \rExact  & 76.5     & 6     & \ttt{while-cubic-T.c} & \rExact  & 84.4     & 7  \\ 
&&&&&                                            \ttt{while-F.c} & \rExact  & 190.4    & 6 \\        
&&&&&                                            \ttt{while-T.c} & \rExact  & 189.6    & 6 \\        
\bottomrule
  \end{tabular}  
\end{table}

Table~\ref{tab:res1}
shows the results of applying \Tool{} to the \benchsvcomp{}, \benchpldi{} and \benchcustom{} benchmarks, respectively. We report whether the final result was validated by Ultimate (\rExact) or needed manual validation (\rAppx). We also report programs that \tool{} cannot handle (marked as \rUNK, \eg~because they use \texttt{double/unsigned} variable types or because they may contain NLA expressions in assignments). Typically refinement completed in around 10 minutes, with most time spent in Ultimate. We also report the number of refinement iterations that were needed ({\bf It.}). We use Ultimate because it appears to be effective in validating the equivalence between NLA and LIA conditions. However, we note that Ultimate cannot discover candidate LIA conditions and does not support CTL whatsoever. Our general strategy is thus to use Ultimate as an ingredient in LIA synthesis, so that our NLA to LIA rewrite can enable NLA support in a broad range of applications, such as CTL for which LIA validation can be done with other tools that support CTL of only LIA programs.

As can be seen, for most programs, \tool{} was able to synthesize correct LIA replacements and often validate them with Ultimate. Even in the cases that need manual validation, the synthesized LIA expressions are correct and, as will be shown in \S\ref{sec:improvements}, can help existing CTL tools.
Morever, \tool{} requires few refinement iterations (usually 2 or fewer and occasionally 4 or 5) to synthesize LIA expressions for \benchsvcomp{} and \benchpldi{}.
\tool{} required more iterations for the \benchcustom{} benchmarks as the LIA expressions for these programs are more complicated.
The following are a few representative examples of NLA expressions in the benchmarks and  LIA expressions generated by \tool{}. The complete output is given in Apx.~\ref{apx:results}.

\begin{center}
  Example output of \Tool{} on \benchsvcomp\\
  \small
  \begin{tabular}{l|l|ll}
    \toprule    
Benchmark & Source NLA & Output $\synthposb$ & Output $\synthnegb$\\
\midrule
\ttt{bresenham1-T.c} &  $\ell_{36}:2Yx-2X^2y+2Y-v+c \leq k$ & $0 \geq c-k$, & $k-c \leq -1$  \\
\ttt{cohencu2-T.c} & $\ell_{32}:3n^2 + 3n + 1 \leq k$ & $0 \geq y - k$, & $k-y \leq -1$ \\
\ttt{egcd2-T.c} & $\ell_{33}: c \geq xq + ys$ & $0 \geq b-c$, & $-b + c \leq -1 $  \\
\bottomrule
\end{tabular}
\end{center}
In these benchmarks, within the respective program context, outputs $\synthposb$ and $\synthnegb$ are simpler linear constraints, because some program variables in the source NLA sum to 0, and therefore their loop conditions can be equally captured by our outputs that involve fewer program variables.
In the case of \ttt{bresenham1-T.c}, Ultimate was able to validate the final $\synthposb$ /$\synthnegb$ pair. In the other cases, Ultimate timed out, so we verified them manually (like all \rAppx{} examples). While manual work is required in these cases, we believe that tools such as Ultimate will continue to improve their support for reachability of NLA programs, and if we accept their current, temporary limitations, we can tackle problems beyond reachability
(such as CTL+NLA verification discussed next, for which no tools exist).

\begin{center}
  Example output of \Tool{} on \benchpldi\\
  \small
\begin{tabular}{l|l|ll}
    \toprule
Benchmark & Source NLA & Output $\synthposb$ & Output $\synthnegb$\\
\midrule
\ttt{afefp-T.c} &  $\ell_{19}:t^2-4s+2t+1+c \leq k$ & $0 \geq a-k$ & $k-a\leq -1$  \\
\ttt{afagp-T.c} &  $\ell_{21}:\neg (xz -x -y +1 +c < k)$ & $0 \geq -c +k$&$c-k \leq -1$  \\
\ttt{neg-afp-F.c} & $\ell_{18}:z^2-12y-6z+12+c \leq k$ &$0 \geq n-k$ & $k-n \leq -1$ \\
\bottomrule
\end{tabular}
\end{center}
The above table provides further examples, this time for the \benchpldi{} benchmarks. Here too, the output LIA expressions are far simpler than the input NLA expressions (\eg~using fewer variables) because \tool{} only involves variables that matter to the semantics of the loop and ignore those that do not (\eg~variables that sum up to zero).

The handcrafted example \ttt{if-cubic-F.c} contained $8 = x^3$ and we 
synthesized the following:
\[\footnotesize
\begin{array}{ll}
\synthposb: &  [(4 \ge p + x) \wedge  (0 \ge p - x) \wedge  (0 \ge -p + x)  \wedge  (-p - x \le -4)] 
 \vee [2 \ge p \wedge  (0 \ge -p + x) \wedge  (-p - x \le -4)]\\
\synthnegb: &  (0 \ge x) \vee [(0 == p - 2 \wedge  1 \ge p - x 
 \;\;\wedge  \neg((2 \ge p \wedge  (0 \ge -p + x) \wedge  (-p - x \le -4)))) \\
\end{array}
\]
%
%
%

Given the program context that $p=2$, $\synthposb$ is equivalent to $8=x^3$, which is validated by Ultimate, and a CTL property can now be verified for this NLA program.
In short, \Tool{} is indeed able to synthesize LIA alternatives for NLA expressions across a variety of benchmarks.

Note although we exploit DIG to generate candidate NLA expressions, 
DIG of its own accord cannot
generate \emph{sound} replacements, nor does DIG have any knowledge of 
properties (beyond guessing candidate invariants) such as termination, temporal logic, slicing, etc.
Instead DIG merely guesses candidate invariants based on sample data. 
Therefore it must be integrated into our iterative refinement loop (and indeed, multiple iterations are often needed for rewriting).

\paragraph{Performance of re-written program}
A natural question is the execution speed of the transformed program compared to the original. To this end, we performed a simple experiment, executing both the original and transformed program on the SVCOMP benchmarks {\benchsvcomp}. For each program pair, we generated 50 random input sets, fed them to each program and measured the execution times. We found that re-writing did not change execution time very much: in essentially all cases, the median execution was within a 4\% difference in time
(sometimes in favor of the re-writtten and sometimes in favor of the original).
We believe this is because the main overhead is setting up program execution, and there is a fairly negligible difference between a few arithmetic operations, many of which may be optimized by the compiler in either case.

\subsection{Enabling CTL Verification of NLA Programs}\label{sec:improvements}

\begin{table}
  \footnotesize  
  \caption{\tool{} enabling T2 and \FuncTion{} on \benchsvcomp{}, \benchpldi{} and \benchcustom{}. The result of applying the tools on the original programs as well as on the rewritten ones by \Tool{}, is shown.}\label{tab:res4}
  \renewcommand{\arraystretch}{\benchtablerowstretch}\setlength{\tabcolsep}{\benchtabletabcolsep}  
  \begin{tabular}{l|crcr|crcr||l|crcr|crcr}
    \toprule
  & \multicolumn{4}{c|}{Enabling T2}   & \multicolumn{4}{c||}{Enabling \FuncTion} & &\multicolumn{4}{c|}{Enabling T2}   & \multicolumn{4}{c}{Enabling \FuncTion}  \\
  & \multicolumn{2}{c}{T2}  & \multicolumn{2}{c|}{\Tool}   & \multicolumn{2}{c}{FT}  & \multicolumn{2}{c||}{\Tool}  & & \multicolumn{2}{c}{T2}  & \multicolumn{2}{c|}{\Tool}   & \multicolumn{2}{c}{FT}  & \multicolumn{2}{c}{\Tool}   \\
Benchmark  & Res & T(s) &  Res & T(s) &  Res &  T(s) &  Res &  T(s) & Benchmark  & Res & T(s) &  Res & T(s) &  Res &  T(s) &  Res &  T(s) \\
  \midrule
\ttt{bresenham1-F.c} & \unsound{\rTRUE} & 0.7      & \rCRASH  & 1.7      & \rUNK    & 5.9      & \rUNK    & 10.9        & \ttt{fermat1-F.c} & \unsound{\rTRUE} & 2.1      & -        & -        & \rUNK    & 1.3      & -        & -         \\       
\ttt{bresenham1-T.c} & \rTRUE   & 0.7      & \rTRUE   & 0.7      & \rUNK    & 0.6      & \rUNK    & 2.7                 & \ttt{fermat1-T.c} & \rTRUE   & 1.9      & -        & -        & \rUNK    & 0.4      & -        & -         \\               
\ttt{cohencu2-F.c} & \unsound{\rTRUE} & 0.7      & \hlg \rFALSE & 0.8      & \rUNK    & 4.4      & \rUNK    & 0.2       & \ttt{geo1-F.c}  & \unsound{\rTRUE} & 1.3      & \hlg \rFALSE & 1.4      & \rUNK    & 0.7      & \rUNK    & 1.8       \\     
\ttt{cohencu2-T.c} & \unsound{\rTRUE}   & 0.7      & \hlg \rTRUE & 0.8      & \rUNK    & 0.1      & \rUNK    & 0.7      & \ttt{geo1-T.c}  & \unsound{\rTRUE}   & 1.4      & \hlg \rTRUE & 1.6      & \rUNK    & 0.1      & \hlg \rTRUE & 0.6       \\           
\ttt{cohencu3-F.c} & \unsound{\rTRUE} & 0.7      & \hlg \rFALSE & 0.8      & \rUNK    & 0.1      & \rUNK    & 0.7       & \ttt{geo2-F.c}  & \unsound{\rTRUE} & 1.5      & \hlg \rFALSE & 1.6      & \rUNK    & 0.7      & \rUNK    & 0.5       \\     
\ttt{cohencu3-T.c} & \unsound{\rTRUE}   & 0.7      & \hlg \rTRUE & 0.8      & \rUNK    & 0.1      & \rUNK    & 0.7      & \ttt{geo2-T.c}  & \unsound{\rTRUE}   & 1.5      & \hlg \rTRUE & 1.4      & \rUNK    & 0.1      & \hlg \rTRUE & 0.6       \\           
\ttt{cohencu4-F.c} & \unsound{\rTRUE} & 0.7      & \hlg \rFALSE & 0.7      & \rUNK    & 0.0      & \rUNK    & 0.0       & \ttt{geo3-F.c}  & \unsound{\rTRUE} & 1.5      & \hlg \rFALSE & 1.4      & \rUNK    & 0.8      & \rUNK    & 4.6       \\     
\ttt{cohencu4-T.c} & \unsound{\rTRUE}   & 0.7      & \hlg \rTRUE & 0.7      & \rUNK    & 0.2      & \hlg \rTRUE & 0.7             & \ttt{geo3-T.c}  & \unsound{\rTRUE}   & 1.5      & \hlg \rTRUE & 1.4      & \rUNK    & 0.1      & \hlg \rTRUE & 0.7       \\           
\ttt{cohencu5-F.c} & \unsound{\rTRUE} & 0.7      & \hlg \rFALSE & 0.7      & \unsound{\rTRUE} & 1.1      & \rUNK  & 0.5 & \ttt{hard-F.c}  & \unsound{\rTRUE} & 1.5      & -        & -        & \rUNK    & 1.6      & -        & -         \\         
\ttt{cohencu5-T.c} & \unsound{\rTRUE}   & 0.7      & \hlg \rTRUE & 0.7      & \rUNK    & 0.2      & \hlg \rTRUE & 1.0             & \ttt{hard-T.c}  & \rTRUE   & 1.6      & -        & -        & \rUNK    & 5.3      & -        & -         \\                 
\ttt{cohencu7-F.c} & \unsound{\rTRUE} & 0.7      & \rCRASH  & 1.7      & \rUNK    & 1.5      & \rUNK    & 0.5           & \ttt{hard2-F.c} & \unsound{\rTRUE} & 1.4      & -        & -        & \rUNK    & 0.0      & -        & -         \\         
\ttt{cohencu7-T.c} & \rTRUE   & 0.8      & \rTRUE   & 0.7      & \rUNK    & 0.4      & \hlg \rTRUE & 1.3                & \ttt{hard2-T.c} & \rTRUE   & 1.5      & -        & -        & \rUNK    & 0.0      & -        & -         \\                 
\ttt{dijkstra2-F.c} & \unsound{\rTRUE} & 0.8      & \unsound{\rTRUE} & 0.8      & \rUNK    & 0.0      & \rUNK    & 0.0  & \ttt{prod4br-F.c} & \unsound{\rTRUE} & 2.0      & -        & -        & \rUNK    & 2.0      & -        & -         \\       
\ttt{dijkstra2-T.c} & \rTRUE   & 0.8      & \rTRUE   & 0.7      & \rUNK    & 670.1    & \rUNK    & \rTO                 & \ttt{prod4br-T.c} & \rCRASH  & 2.4      & -        & -        & \rUNK    & 0.5      & -        & -         \\               
\ttt{dijkstra3-F.c} & \unsound{\rTRUE} & 0.8      & \unsound{\rTRUE} & 0.8      & \rUNK    & \rTO     & \rUNK    & \rTO & \ttt{prodbin-F.c} & \unsound{\rTRUE} & 1.3      & -        & -        & \rUNK    & 0.5      & -        & -         \\       
\ttt{dijkstra3-T.c} & \rCRASH  & 1.7      & \rCRASH  & 1.7      & \rUNK    & 642.9    & \rUNK    & \rTO                 & \ttt{prodbin-T.c} & \rTRUE   & 1.1      & -        & -        & \rUNK    & 0.2      & -        & -         \\               
\ttt{dijkstra4-F.c} & \unsound{\rTRUE} & 0.8      & \rCRASH  & 1.8      & \rUNK    & \rTO     & \rUNK    & \rTO         & \ttt{ps2-F.c}   & \unsound{\rTRUE} & 1.0      & \hlg \rFALSE & 1.0      & \rUNK    & 0.6      & \rUNK    & 1.5       \\     
\ttt{dijkstra4-T.c} & \rTRUE   & 0.8      & \rTRUE   & 0.7      & \rUNK    & 603.8    & \rUNK    & \rTO                 & \ttt{ps2-T.c}   & \unsound{\rTRUE}   & 0.6      & \hlg \rTRUE & 0.7      & \rUNK    & 0.1      & \hlg \rTRUE & 0.5       \\           
\ttt{dijkstra5-F.c} & \unsound{\rTRUE} & 0.8      & \unsound{\rTRUE} & 0.8      & \rUNK    & \rTO     & \rUNK    & \rTO & \ttt{ps3-F.c}   & \unsound{\rTRUE} & 0.7      & \hlg \rFALSE & 0.7      & \rUNK    & 0.6      & \rUNK    & 1.5       \\     
\ttt{dijkstra5-T.c} & \rTRUE   & 0.8      & \rTRUE   & 0.7      & \rUNK    & 738.5    & \rUNK    & \rTO                 & \ttt{ps3-T.c}   & \unsound{\rTRUE}   & 0.7      & \hlg \rTRUE & 1.1      & \rUNK    & 0.1      & \hlg \rTRUE & 0.5       \\           
\ttt{divbin1-F.c} & \rUNK    & 0.2      & -        & -        & \rUNK    & 0.0      & -        & -                      & \ttt{ps4-F.c}   & \unsound{\rTRUE} & 0.7      & \hlg \rFALSE & 0.7      & \rUNK    & 0.6      & \rUNK    & 1.5       \\     
\ttt{divbin1-T.c} & \rUNK    & 0.2      & -        & -        & \rUNK    & 0.0      & -        & -                      & \ttt{ps4-T.c}   & \unsound{\rTRUE}   & 0.7      & \hlg \rTRUE & 0.7      & \rUNK    & 0.1      & \hlg \rTRUE & 0.5       \\           
\ttt{egcd-F.c}  & \unsound{\rTRUE} & 0.8      & -        & -        & \rUNK    & 0.2      & -        & -                & \ttt{ps5-F.c}   & \unsound{\rTRUE} & 0.7      & \hlg \rFALSE & 0.7      & \rUNK    & 0.6      & \rUNK    & 1.5       \\     
\ttt{egcd-T.c}  & \rCRASH  & 1.8      & -        & -        & \rUNK    & 0.2      & -        & -                        & \ttt{ps5-T.c}   & \unsound{\rTRUE}   & 0.9      & \hlg \rTRUE & 0.9      & \rUNK    & 0.1      & \hlg \rTRUE & 0.5       \\           
\ttt{egcd2-F.c} & \unsound{\rTRUE} & 1.9      & \unsound{\rTRUE} & 1.4      & \rUNK    & 3.1      & \rUNK    & 6.3      & \ttt{ps6-F.c}   & \unsound{\rTRUE} & 1.0      & \hlg \rFALSE & 0.9      & \rUNK    & 0.6      & \rUNK    & 1.5       \\     
\ttt{egcd2-T.c} & \rTRUE   & 1.3      & \rTRUE   & 1.6      & \rUNK    & 1.7      & \rUNK    & 11.9                     & \ttt{ps6-T.c}   & \unsound{\rTRUE}   & 1.0      & \hlg \rTRUE & 1.0      & \rUNK    & 0.1      & \hlg \rTRUE & 0.5       \\           
\ttt{egcd3-F.c} & \unsound{\rTRUE} & 1.5      & -        & -        & \rUNK    & 0.0      & -        & -                & \ttt{sqrt1-F.c} & \unsound{\rTRUE} & 0.8      & \hlg \rFALSE & 0.8      & \unsound{\rTRUE} & 0.8      & \rUNK    & 2.3 \\   
  \ttt{egcd3-T.c} & \rTRUE   & 1.8      & -        & -        & \rUNK    & 18.9     & -        & -                      & \ttt{sqrt1-T.c} & \unsound{\rTRUE}   & 0.7      & \hlg \rTRUE & 0.6      & \rUNK    & 0.1      & \hlg \rTRUE & 0.8                  \\
    \bottomrule
  \end{tabular}  

  \renewcommand{\arraystretch}{\benchtablerowstretch}\setlength{\tabcolsep}{\benchtabletabcolsep}  
  \begin{tabular}{l|crcr|crcr||l|crcr|crcr}
    \toprule
  & \multicolumn{4}{c|}{Enabling T2}   & \multicolumn{4}{c||}{Enabling \FuncTion} & &\multicolumn{4}{c|}{Enabling T2}   & \multicolumn{4}{c}{Enabling \FuncTion}  \\
  & \multicolumn{2}{c}{T2}  & \multicolumn{2}{c|}{\Tool}   & \multicolumn{2}{c}{FT}  & \multicolumn{2}{c||}{\Tool}  & & \multicolumn{2}{c}{T2}  & \multicolumn{2}{c|}{\Tool}   & \multicolumn{2}{c}{FT}  & \multicolumn{2}{c}{\Tool}   \\
  Benchmark  & Res & T(s) &  Res & T(s) &  Res &  T(s) &  Res &  T(s) & Benchmark  & Res & T(s) &  Res & T(s) &  Res &  T(s) &  Res &  T(s) \\
  \midrule
\ttt{afagp-F.c} & \rFALSE  & 1.6      & \rFALSE  & 1.9      & \rUNK    & 0.0      & \rUNK    & 0.0             &\ttt{afp-F.c}   & \rFALSE  & 1.0      & \rFALSE  & 1.1      & \rUNK    & 0.0      & \unsound{\rTRUE} & 0.1       \\       
\ttt{afagp-T.c} & \unsound{\rFALSE} & 1.2      & \hlg \rTRUE & 1.2      & \rUNK    & 0.0      & \rUNK    & 0.2 & \ttt{afp-T.c}   & \rTRUE   & 0.9      & \rTRUE   & 1.2      & \rTRUE   & 0.0      & \rTRUE   & 0.1       \\               
\ttt{afefp-T.c} & \unsound{\rFALSE} & 16.6     & \hlg \rTRUE & 25.9     & \rUNK    & 0.0      & \rUNK    & 0.0 & \ttt{agafp-F.c} & \rFALSE  & 2.2      & \rFALSE  & 1.6      & \rUNK    & 0.1      & \rUNK    & 0.6       \\               
\ttt{afegp-F.c} & \rFALSE  & 1.8      & \rFALSE  & 1.8      & \rUNK    & 0.1      & \rUNK    & 0.0             & \ttt{agafp-T.c} & \unsound{\rFALSE} & 1.3      & \hlg \rTRUE & 1.0      & \rUNK    & 0.0      & \rUNK    & 1.0       \\   
  \ttt{afegp-T.c} & \rTRUE   & 1.7      & \rTRUE   & 1.8      & \rUNK    & 0.1      & \rUNK    & 0.2             & \ttt{efafp-T.c} & \rTRUE   & 3.4      & \rTRUE   & 2.1      & \rUNK    & 0.2      & \rUNK    & 1.5       \\
\ttt{neg-afagp-F.c} & \rFALSE  & 1.4      & \rFALSE  & 1.5      & \unsound{\rTRUE} & 0.1      & \rUNK    & 0.2       &         \ttt{neg-afp-F.c} & \unsound{\rTRUE} & 1.1      & \hlg \rFALSE & 1.0      & \rUNK    & 0.0      & \rUNK    & 0.1       \\           
\ttt{neg-afagp-T.c} & \rTRUE   & 2.1      & \rTRUE   & 2.1      & \rTRUE   & 0.1      & \rUNK    & 0.2               &         \ttt{neg-afp-T.c} & \unsound{\rTRUE}   & 1.1      & \hlg \rTRUE & 1.1      & \rUNK    & 0.0      & \rUNK    & 0.1       \\                    
\ttt{neg-afefp-F.c} & \unsound{\rTRUE} & 13.3     & \hlg \rFALSE & 57.1     & \rUNK    & 0.2      & \rUNK    & 0.7   &              \ttt{neg-efafp-F.c} & \rFALSE  & 3.8      & \rFALSE  & 2.4      & \rUNK    & 0.1      & \rUNK    & 0.4       \\                     
\ttt{neg-afegp-F.c} & \unsound{\rTRUE} & 1.7      & \unsound{\rTRUE} & 1.7      & \unsound{\rTRUE} & 0.1      & \rUNK    & 0.2    &  \ttt{neg-egafp-F.c} & \rFALSE  & 2.1      & \rFALSE  & 1.4      & \rUNK    & 0.0      & \rUNK    & 0.2       \\                     
\ttt{neg-afegp-T.c} & \unsound{\rFALSE} & 1.6      & \unsound{\rFALSE} & 1.7      & \rTRUE   & 0.1      & \rUNK    & 0.2          &       \ttt{neg-egafp-T.c} & \unsound{\rFALSE} & 2.3      & \unsound{\rFALSE} & 1.6      & \rUNK    & 0.0      & \rUNK    & 0.2       \\   
\ttt{neg-egimpafp-T.c} & \unsound{\rFALSE} & 2.4      & \unsound{\rFALSE} & 1.7      & \rUNK    & 0.1      & \rUNK    & 0.6    &&&&&&&&&   \\
  \midrule
\ttt{afagp-T.c} & \unsound{\rFALSE} & 1.3      & \unsound{\rFALSE} & 1.2      & \rUNK    & 0.0      & \rUNK    & 0.2   & \ttt{neg-afefp-F.c} & \unsound{\rTRUE} & 13.2     & \hlg \rFALSE & 22.3     & \rUNK    & 0.1      & \rUNK    & 0.8     \\
\ttt{afefp-T.c} & \unsound{\rFALSE} & 12.9     & \hlg \rTRUE & 25.7     & \rUNK    & 0.7      & \rUNK    & 2.4       &   \ttt{neg-afp-F.c} & \unsound{\rTRUE} & 1.0      & \hlg \rFALSE & 0.9      & \rUNK    & 0.0      & \rUNK    & 0.1       \\
\ttt{agafp-T.c} & \unsound{\rFALSE} & 1.2      & \hlg \rTRUE & 0.9      & \rUNK    & 0.0      & \rUNK    & 0.5       &&&&&&&\\
\bottomrule
  \end{tabular}  

  \renewcommand{\arraystretch}{\benchtablerowstretch}\setlength{\tabcolsep}{\benchtabletabcolsep}  
  \begin{tabular}{l|crcr|crcr||l|crcr|crcr}
    \toprule
  & \multicolumn{4}{c|}{Enabling T2}   & \multicolumn{4}{c||}{Enabling \FuncTion} & &\multicolumn{4}{c|}{Enabling T2}   & \multicolumn{4}{c}{Enabling \FuncTion}  \\
  & \multicolumn{2}{c}{T2}  & \multicolumn{2}{c|}{\Tool}   & \multicolumn{2}{c}{FT}  & \multicolumn{2}{c||}{\Tool}  & & \multicolumn{2}{c}{T2}  & \multicolumn{2}{c|}{\Tool}   & \multicolumn{2}{c}{FT}  & \multicolumn{2}{c}{\Tool}   \\
  Benchmark  & Res & T(s) &  Res & T(s) &  Res &  T(s) &  Res &  T(s) & Benchmark  & Res & T(s) &  Res & T(s) &  Res &  T(s) &  Res &  T(s) \\
  \midrule
\ttt{if-cubic-F.c} & \unsound{\rTRUE} & 0.9      & \hlg \rFALSE & 0.7      & \rUNK    & 0.1      & \rUNK    & 0.1  &  \ttt{square-loop-F.c} & \rFALSE  & 0.8      & \unsound{\rTRUE} & 0.8      & \rUNK    & 0.0      & \rUNK    & 0.0       \\         
\ttt{if-cubic-T.c} & \unsound{\rTRUE}   & 0.6      & \hlg \rTRUE & 0.7      & \rUNK    & 0.0      & \rUNK    & 0.0           &           \ttt{square-loop-T.c} & \rTRUE   & 0.8      & \unsound{\rFALSE} & 0.9      & \rUNK    & 0.1      & \rUNK    & 3.1       \\        
\ttt{if-F.c}    & \unsound{\rTRUE} & 0.7      & \hlg \rFALSE & 0.7      & \rUNK    & 0.0      & \rUNK    & 0.1     &     \ttt{while-cubic-F.c} & \unsound{\rTRUE} & 0.7      & \hlg \rFALSE & 0.8      & \rUNK    & 0.0      & \rUNK    & 0.2       \\     
\ttt{if-T.c}    & \unsound{\rTRUE}   & 0.7      & \hlg \rTRUE & 0.7      & \rUNK    & 0.0      & \rUNK    & 0.1    &      \ttt{while-cubic-T.c} & \unsound{\rTRUE}   & 0.7      & \hlg \rTRUE & 0.7      & \rUNK    & 0.0      & \rUNK    & 0.5       \\              
&&&&&&&&&                                                                                                                            \ttt{while-F.c} & \rFALSE  & 0.9      & \rFALSE  & 0.9      & \rUNK    & 0.1      & \rUNK    & 0.4       \\                       
&&&&&&&&&                                                                                                                            \ttt{while-T.c} & \rTRUE   & 0.7      & \rTRUE   & 0.8      & \rUNK    & 0.1      & \rUNK    & 3.2       \\                       
\bottomrule
  \end{tabular}  
\end{table}

Table~\ref{tab:res4}
shows how \tool{} enables the two state of the art CTL verifiers T2~\cite{Cook2015} and \FuncTion~\cite{function} to support NLA programs. 
In this table, under the Enabling T2 (resp., \FuncTion) group, columns \textbf{Res} under T2 (\FuncTion) and \tool{} show the results of running T2 (\FuncTion) on the original program and on the \tool{}'s generated program, respectively.   The symbol \rTRUE means proved (expected for benchmarks with T-suffix), \rFALSE{} means disproved (expected for benchmarks with F-suffix), those with \unsound{} means incorrect result (e.g., \unsound{\rTRUE} is unsound as it proves an invalid property), \rCRASH{} means crash, \rUNK{} means unknown and `-' means \tool{} was not able to rewrite NLA expressions. Results with \colorbox{green!20}{green background} 
indicate cases where thanks to \tool{}, the CTL verifier was able to verify the NLA program when it previously was not able to do so.

(1) {\benchsvcomp.} As can be seen from the first stanza of Table~\ref{tab:res4}, T2 did not support programs with NLAs (it incorrectly returned `Valid' even on invalid examples, \ie~emitting unsound output).
However, with \tool{}'s help, T2 was able to correctly verify 26/56 programs (green background).
For the other 30/56 programs that \tool{} failed to help, it is because \tool{} was not able to rewrite 14 programs (due to NLAs in assignments, Double/Unsigned variable types etc.), because for 13 programs that \tool{} could rewrite, T2 could still not verify them correctly, and because T2 would still crash on 3 of \tool{}'s generated programs (arguably still better than the unsound results that T2 had for these programs).

\FuncTion{} failed to run for all but two programs, \texttt{cohencu5-T} and \texttt{sqrt1-F}, in which it ran but gave unsound results, proving invalid properties.
With \tool{}'s help, \FuncTion{} was able to correctly analyze 12/56 programs.
The running times of T2 and \FuncTion{} with the original programs and \tool{}'s generated programs are similar (typically $<1$ second). This is expected as \tool{}'s generated programs are simpler than the original, and should not cost extra analysis time.

(2) {\benchpldi.} As seen in the second stanza of Table~\ref{tab:res4}, T2 also had difficulty analyzing the instrumented NLA code and returned unsound results. \FuncTion{} did not run on most programs, and in the few cases that it did, it likely gave unsound results.  \tool{} enabled T2 to verify 10 benchmarks,
but did not appear to help \FuncTion. Differences in running times are negligible.

(3) {\benchcustom.} In the final stanza of Table~\ref{tab:res4}, as expected, both T2 and \FuncTion{} had difficulties and returned incorrect results for some programs. Here, \tool{} could not help \FuncTion{} but could help T2 verify 6 benchmarks. However, it appears that \tool{}'s generated  \texttt{square-loop} programs with complex linear constraints confused T2 and caused it to return incorrect results. 

In summary, for NLA benchmarks, T2 and \FuncTion{} simply do not support these programs: 
T2 appears it shouldn't be run on them (it emits ``valid'' on invalid examples) and \FuncTion{} always returns ``unknown.''
With the help of \tool{}, these tools can now analyze many programs (42 out of 92) that they previously could not.
Moreover, if one were to use \tool{} as a pre-processing to enable existing tools to support NLA, then these improvements come with almost no additional runtime cost.
To be clear: the pre-processing of \tool{} may take a few minutes, but then the CTL verification speed
(on these NLA programs that previously could not be handled) is the fast, typical speed of those CTL tools on LIA programs  (in most cases the analyses took less than a second).


\subsection{Improving Program Slicing }\label{sec:slicing}

Static program slicing tools often struggle when there are NLA expressions along the slicing paths, unable to slice away code that is independent from the slicing criteria. 
By first applying dual rewriting, we can replace these troublesome expressions with linear alternatives that are easier for those tools to reason about. We now give a simple proof-of-concept example and show that dual re-writing enables Frama-C to slice a program it otherwise could not. In future work dual rewriting could be integrated more deeply into compilation pipelines, upstream of tools like Frama-C.

\begin{figure}[!h]
\begin{quote}
\begin{lstlisting}[language=Python,style=pnnstyle]
int X, Y, v, x, y;
int X = read();
int Y = read();
int v = 2 * Y - X;
int y=0, x=0, c=0, k=0, p=2;
while(2*Y*x-2*X*y-X+2*Y-v+c<=k-1){ // |\boxed{\text{\Tool{}}: c <= k-1}\label{ln:frama-loop}|
  if (v < 0) {
       v = v + 2 * Y;
    } else {
       v = v + 2 * (Y - X);
       y++;
    }
      x++;
      c++;
      p = 1;
  }
if (p != 1) print(p); |\label{ln:frama-if}|
return ; \end{lstlisting}
\end{quote}
\caption{\label{fig:slice-program} Program slicing examples with NLA expression.}
\end{figure}

Figure~\ref{fig:slice-program} lists a program with a polynomial guard in the \lstinline|while| loop. After the loop the function \lstinline|print(p)| is called provided that \lstinline|p != 1|. 
Out of the box, asking Frama-C to slice to the target \lstinline|print(p)| yields an unchanged output program, i.e., it cannot remove any statements.

By contrast, without any changes to Frama-C and merely using \Tool{} as a pre-processing step, we were able to drastically slice this program down to the small program to the right.
\begin{wrapfigure}[5]{r}[0pt]{1.2in}
\begin{lstlisting}[language=Python,style=pnnstyle]
  int p;
  p = 2;
  print(p);
  return;
\end{lstlisting}
\end{wrapfigure}
Specifically, we first run \Tool{} which discovers the alternative LIA condition for the while loop  
on Line~\ref{ln:frama-loop} of Fig.~\ref{fig:slice-program}.
We replace the NLA loop condition with this simpler \lstinline|c <= k-1| and then run Frama-C.
Now Frama-C is able to determine that the loop condition is infeasible, so there is no risk that 
the assignment \lstinline|p=1| will execute. Thus the entire loop can be sliced away as well as the conditional
\lstinline|p!=1| on Line~\ref{ln:frama-if}.
Without \Tool{}'s LIA replacement, Frama-C over-approximates the NLA loop guard with the assumption that it could evaluate to true. Thus the assignment \lstinline|p=1| could execute and so conditional
\lstinline|p!=1| on Line~\ref{ln:frama-if} is also necessary. Therefore, Frama-C's slicing cannot remove any of the program.


\section{Conclusion}

We have shown that, for NLA program expressions, semantically equivalent boolean linear arithmetic expressions can be  algorithmically discovered. This enables off-the-shelf verification tools for LIA programs to be applied to a larger range of NLA input programs. We showed this is true for CTL verification tools and some evidence that dual rewriting can impact program slicing.
The outcome is that \Tool{} can be used as a pre-processor for many possible analysis purposes. So a tool that currently does not support NLA expressions can perhaps be easily made to support them by simply pre-running \Tool{}.

In future work we hope to apply dual rewriting to a wider variety of application domains. We will explore integrating dual rewriting into compiler frameworks and explore how our rewriting interacts with other compilation analyses and optimizations. We will so explore application to other automated verification concerns such as fault localization.

\begin{acks} 
This work is supported in part by the 
\grantsponsor{GSNSF}{National Science Foundation}{https://www.nsf.gov}
awards
\grantnum{GSNSF}{CCF-2107169} (Stevens),
\grantnum{GSNSF}{CCF-2106845} (Yale), and
\grantnum{GSNSF}{CCF-2318030} (George Mason).
Yuandong Liu and Eric Koskinen were also supported in part by 
the \grantsponsor{GSONR}{Office of Naval Research}{https://www.nre.navy.mil/} 
under Grant No.~\grantnum{GSONR}{N00014-22-1-2643}.
\end{acks}


\pagebreak
\appendix
\section{CTL Semantics}
\label{apx:ctlsem}

\begin{displaymath}
\begin{array}{rll}
M, \sigma \vDash p & \Longleftrightarrow & \sigma \in \sem{p} \\ 
M, \sigma \vDash \prop_{1} \wedge \prop_{2} & \Longleftrightarrow & M, \sigma \vDash \prop_{1} \wedge M, \sigma \vDash \prop_{2} \\ 
M, \sigma \vDash \prop_{1} \vee \prop_{2} & \Longleftrightarrow & M, \sigma \vDash \prop_{1} \vee M, \sigma \vDash \prop_{2} \\ 
M, \sigma \vDash \AF \prop & \Longleftrightarrow & \forall (\sigma_{0}, \sigma_{1}, ...) \in \Pi(\states, R, \{\sigma\}). ~\exists i \ge 0. ~M, \sigma_{i} \vDash \prop \\ 
M, \sigma \vDash \EF \prop & \Longleftrightarrow & \exists (\sigma_{0}, \sigma_{1}, ...) \in \Pi(\states, R, \{\sigma\}). ~\exists i \ge 0. ~M, \sigma_{i} \vDash \prop \\ 
M, \sigma \vDash \A \lbrack \prop_{1} \WW \prop_{2} \rbrack & \Longleftrightarrow &
                    \forall (\sigma_{0}, \sigma_{1}, ...) \in \Pi(\states, R, \{\sigma\}). ~(\forall i \ge 0. ~M, \sigma_{i} \vDash \prop_{1}) \vee\\
                 &&\hspace{50pt}   (\exists j \ge 0. ~M,\sigma_{j} \vDash \prop_{2} \wedge \forall i \in \lbrack 0, j). ~M, \sigma_{i} \vDash \prop_{1}) \\ 
M, \sigma \vDash \EE \lbrack \prop_{1} \WW \prop_{2} \rbrack & \Longleftrightarrow &
                    \exists (\sigma_{0}, \sigma_{1}, ...) \in \Pi(\states, R, \{\sigma\}). ~(\forall i \ge 0. M, \sigma_{i} \vDash \prop_{1}) \vee\\
					&&\hspace{50pt} (\exists j \ge 0. ~M,\sigma_{j} \vDash \prop_{2} \wedge \forall i \in \lbrack 0, j). ~M, \sigma_{i} \vDash \prop_{1}) 
\end{array}
\end{displaymath}

\section{Proof of Lemma~\ref{lemma:sound}}
\label{apx:soundness}

The following is a sketch of the proof of Lemma~\ref{lemma:sound}.
\begin{proof}
	Consider a mapping $\loc \mapsto (\synthposbi, \synthnegbi) \in m$. Because $\errorPosTooS^\loc$, $\errorPosTooB^\loc$, $\errorNegTooS^\loc$, $\errorNegTooB^\loc$ are unreachable, their corresponding conditional conditions are unsatisfiable under the program context in which $\bloc$ could be evaluated (to either \ttt{true} or \ttt{false}). 
	Therefore, 
	\[\begin{array}[t]{ll}
	& \forall \tstate{} \in \preds{\bloc}.\, \neg(\sem{\bloc}\tstate{} \wedge \neg \sem{\synthposbi}\tstate{}) \wedge \neg(\neg \sem{\bloc}\tstate{} \wedge \sem{\synthposbi}\tstate{})\\
	\Leftrightarrow & \forall \tstate{} \in \preds{\bloc}.\, (\neg \sem{\bloc}\tstate{} \vee \sem{\synthposbi}\tstate{}) \wedge (\sem{\bloc}\tstate{} \vee \neg \sem{\synthposbi}\tstate{})\\
	\Leftrightarrow & \forall \tstate{} \in \preds{\bloc}.\, (\sem{\bloc}\tstate{} \implies \sem{\synthposbi}\tstate{}) \wedge (\sem{\synthposbi}\tstate{} \implies \sem{\bloc}\tstate{})\\
	\Leftrightarrow & \forall \tstate{} \in \preds{\bloc}.\, \sem{\bloc}\tstate{} = \sem{\synthposbi}\tstate{}\\
	\end{array}\]
	Similarly, we can prove that $\forall \tstate{} \in \preds{\bloc}.\, \sem{\neg \bloc}\tstate{} = \sem{\synthnegbi}\tstate{}$.
\end{proof}

\section{Unabridged Experimental Results}
\label{apx:results}

\subsection{\tool{} on \benchsvcomp{} benchmarks}
{\footnotesize
\renewcommand{\arraystretch}{\benchtablerowstretch}\setlength{\tabcolsep}{\benchtabletabcolsep}
\begin{tabular}{|l|c|r|r|p{0.8in}|p{2.8in}|}
\hline
\textbf{Benchmark} & \textbf{Res} & \textbf{Time} & \textbf{It.} & \textbf{Ref. Stages} &
\textbf{Output}  \\
\ttt{bresenham1-F.c} & \rExact  & 210.1    & 2  & v,en,v,v & $\ell_{36}:2Yx-2X^2y+2Y-v+c \leq k$ $\mapsto$ $0 \geq c-k,k-x \leq -1$  \\
\ttt{bresenham1-T.c} & \rExact  & 204.2    & 2  & v,en,v,v & $\ell_{36}:2Yx-2X^2y+2Y-v+c \leq k$ $\mapsto$ $0 \geq c-k,k-c \leq -1$  \\
\ttt{cohencu2-F.c} & \rAppx   & 517.5    & 1  & v,v & $\ell_{28}:3n^2 + 3n + 1 \leq k$ $\mapsto$ $0 \geq y - k,k-y \leq -1$  \\
\ttt{cohencu2-T.c} & \rAppx   & 473.7    & 1  & v,v & $\ell_{32}:3n^2 + 3n + 1 \leq k$ $\mapsto$ $0 \geq y - k,k-y \leq -1$  \\
\ttt{cohencu3-F.c} & \rAppx   & 621.8    & 1  & v,v & $\ell_{31}:n^3 \leq k$ $\mapsto$ $0 \geq c-k,((k - x) <= -(1))$  \\
\ttt{cohencu3-T.c} & \rAppx   & 621.4    & 1  & v,v & $\ell_{31}:n^3 \leq k$ $\mapsto$ $0 \geq c-k,((k - x) <= -(1))$  \\
\ttt{cohencu4-F.c} & \rAppx   & 780.3    & 2  & v,en,v,v & $\ell_{31}:yz-18x-12y+2z-6+c \leq k$ $\mapsto$ $0 \geq c-k,k-c \leq -1$  \\
\ttt{cohencu4-T.c} & \rAppx   & 773.9    & 2  & v,en,v,v & $\ell_{31}:yz-18x-12y+2z-6+c \leq k$ $\mapsto$ $0 \geq c-k,k-c \leq -1$  \\
\ttt{cohencu5-F.c} & \rExact  & 179.4    & 2  & v,en,v,v & $\ell_{35}:z^2-12y-6z+12+c \leq k$ $\mapsto$ $0 \geq c-k,k-c \leq -1$  \\
\ttt{cohencu5-T.c} & \rExact  & 177.4    & 2  & v,en,v,v & $\ell_{35}:z^2-12y-6z+12+c \leq k$ $\mapsto$ $0 \geq n-k,(((0 + (k   1)) + (n   -1)) <= -1)$  \\
\ttt{cohencu7-F.c} & \rAppx   & 194.9    & 2  & v,en,v,v & $\ell_{30}:((x + y) <= (((a + 1)   (a + 1))   (a + 1)))$ $\mapsto$ $(0 >= (-(a) + n)),(((0 + (n   -1)) + (a   1)) <= -1)$  \\
\ttt{cohencu7-T.c} & \rAppx   & 667.5    & 2  & v,en,v,v & $\ell_{28}:((x + y) <= (((a + 1)   (a + 1))   (a + 1)))$ $\mapsto$ $(0 >= (-(a) + n)),(((0 + (n   -1)) + (a   1)) <= -1)$  \\
\ttt{dijkstra2-F.c} & \rAppx   & 687.5    & 2  & v,en,v,v & $\ell_{46}:(((((xp   xp) + (r   q)) - (n   q)) + c) <= k)$ $\mapsto$ $0 \geq c-k,k-c \leq -1$  \\
\ttt{dijkstra2-T.c} & \rAppx   & 686.0    & 2  & v,en,v,v & $\ell_{48}:(((((xp   xp) + (r   q)) - (n   q)) + c) <= k)$ $\mapsto$ $0 \geq c-k,k-c \leq -1$  \\
\ttt{dijkstra3-F.c} & \rAppx   & 628.6    & 1  & v,v & $\ell_{45}:h^3 - 12hnq + 16nx' - hq^2 - 4x'q^2 + 12hqr - 6x'qr +c \leq k$ $\mapsto$ $0 \geq c-k,((-(c) + k) <= -(1))$  \\
\ttt{dijkstra3-T.c} & \rAppx   & 627.5    & 1  & v,v & $\ell_{45}:h^3 - 12hnq + 16nx' - hq^2 - 4x'q^2 + 12hqr - 6x'qr +c \leq k$ $\mapsto$ $0 \geq c-k,((-(c) + k) <= -(1))$  \\
\ttt{dijkstra4-F.c} & \rAppx   & 629.7    & 1  & v,v & $\ell_{45}:((((((((((((h   h)   n) - (((4   h)   n)   xp)) + ((4   (n   n))   q)) - ((n   q)   q)) - ((h   h)   r)) + (((4   h)   xp)   r)) - (((8   n)   q)   r)) + ((q   q)   r)) + (((4   q)   r)   r)) + c) <= k)$ $\mapsto$ $0 \geq c-k,((-(c) + k) <= -(1))$  \\
\ttt{dijkstra4-T.c} & \rAppx   & 629.1    & 1  & v,v & $\ell_{45}:((((((((((((h   h)   n) - (((4   h)   n)   xp)) + ((4   (n   n))   q)) - ((n   q)   q)) - ((h   h)   r)) + (((4   h)   xp)   r)) - (((8   n)   q)   r)) + ((q   q)   r)) + (((4   q)   r)   r)) + c) <= k)$ $\mapsto$ $0 \geq c-k,((-(c) + k) <= -(1))$  \\
\ttt{dijkstra5-F.c} & \rAppx   & 630.8    & 1  & v,v & $\ell_{46}:(((((((((h   h)   xp) - (((4   h)   n)   q)) + (((4   n)   xp)   q)) - ((xp   q)   q)) + (((4   h)   q)   r)) - (((4   xp)   q)   r)) + c) <= k)$ $\mapsto$ $0 \geq c-k,((-(c) + k) <= -(1))$  \\
\ttt{dijkstra5-T.c} & \rAppx    & 613.9    & 1  & v,v & $\ell_{46}:(((((((((h   h)   xp) - (((4   h)   n)   q)) + (((4   n)   xp)   q)) - ((xp   q)   q)) +     (((4   h)   q)   r)) - (((4   xp)   q)   r)) + c) <= k)$ $\mapsto$ $0 \geq c-k,((-(c) + k) <= -(1))$  \\
\ttt{divbin1-F.c} & \rUNK    & 1.2      & 0  &  &  \\
\ttt{divbin1-T.c} & \rUNK    & 1.3      & 0  &  &  \\
\ttt{egcd-F.c}  & \rAppx    & \rTO     & 0  &  &  \\
\ttt{egcd-T.c}  & \rAppx   & \rTO     & 0  &  &  \\
\ttt{egcd2-F.c} & \rAppx   & 696.0    & 1  & v,v & $\ell_{33}:(c >= ((x   q) + (y   s)))$ $\mapsto$ $(0 >= (b - c)),((-(b) + c) <= -(1))$  \\
\ttt{egcd2-T.c} & \rAppx   & 681.8    & 1  & v,v & $\ell_{33}:(c >= ((x   q) + (y   s)))$ $\mapsto$ $(0 >= (b - c)),((-(b) + c) <= -(1))$  \\
\ttt{egcd3-F.c} & \rAppx    & \rTO      & 8  & v,tn,v,tp,v,tn,v,tp,v,tn,v,tp,v,tn,v,tp,v &  \\
\ttt{egcd3-T.c} & \rAppx   & \rTO     & 8  & v,tn,v,tp,v,tn,v,tp,v,tn,v,tp,v,tn,v,tp,v &  \\
\ttt{fermat1-F.c} & \rAppx   & \rTO     & 0  &  &  \\
\ttt{fermat1-T.c} & \rAppx   & \rTO     & 0  &  &  \\
\ttt{geo1-F.c}  & \rExact  & 131.0    & 1  & v,v & $\ell_{24}:!(((((((x   z) - x) - y) + 1) + c) < k))$ $\mapsto$ $(0 >= (-(c) + k)),((c - k) <= -(1))$  \\
\ttt{geo1-T.c}  & \rExact  & 130.6    & 1  & v,v & $\ell_{24}:!(((((((x   z) - x) - y) + 1) + c) < k))$ $\mapsto$ $(0 >= (-(c) + k)),((c - k) <= -(1))$  \\
\ttt{geo2-F.c}  & \rExact  & 122.5    & 1  & v,v & $\ell_{24}:!((((((1 + (x   z)) - x) - (z   y)) + c) < k))$ $\mapsto$ $(0 >= (-(c) + k)),((c - k) <= -(1))$  \\
\ttt{geo2-T.c}  & \rExact  & 134.7    & 1  & v,v & $\ell_{25}:!((((((1 + (x   z)) - x) - (z   y)) + c) < k))$ $\mapsto$ $(0 >= (-(c) + k)),((c - k) <= -(1))$  \\

\end{tabular}}



\bigskip
{\footnotesize
\renewcommand{\arraystretch}{\benchtablerowstretch}\setlength{\tabcolsep}{\benchtabletabcolsep}
\begin{tabular}{|l|c|r|r|p{0.8in}|p{2.8in}|}
\textbf{Benchmark} & \textbf{Res} & \textbf{Time} & \textbf{It.} & \textbf{Ref. Stages} &
\textbf{Output}  \\

\ttt{geo3-F.c}  & \rExact  & 171.1    & 1  & v,v & $\ell_{25}:!(((((((z   x) - x) + a) - ((a   z)   y)) + c) < k))$ $\mapsto$ $(0 >= (-(c) + k)),((c - k) <= -(1))$  \\
\ttt{geo3-T.c}  & \rExact  & 163.5    & 1  & v,v & $\ell_{25}:!(((((((z   x) - x) + a) - ((a   z)   y)) + c) < k))$ $\mapsto$ $(0 >= (-(c) + k)),((c - k) <= -(1))$  \\
\ttt{hard-F.c}  & \rUNK    & 7.9      & 0  &  &  \\
\ttt{hard-T.c}  & \rUNK    & 7.4      & 0  &  &  \\
\ttt{hard2-F.c} & \rUNK    & 0.9      & 0  &  &  \\
\ttt{hard2-T.c} & \rUNK    & 0.8      & 0  &  &  \\
\ttt{prod4br-F.c} & \rAppx   & 239.0    & 5  & v,ep,v,en,v,tn,v,en,v,tn,v &  \\
\ttt{prod4br-T.c} & \rAppx   & 159.3    & 4  & v,ep,v,en,v,tn,v,en,v &  \\
\ttt{prodbin-F.c} & \rAppx   & 654.5    & 2  & v,en,v,v & $\ell_{26}:!((0 != (((y + z) + (x   y)) - (a   b))))$ $\mapsto$ $(0 >= y),((0 + (y   -1)) <= -1)\&\&(((0 + (p   -1)) + (y   -1)) <= -2)\&\&((0 + (p   -1)) <= -1)\&\&(((0 + (p   1)) + (y   -1)) <= 1)$  \\
\ttt{prodbin-T.c} & \rAppx   & 619.6    & 1  & v,v & $\ell_{26}:!((0 != (((y + z) + (x   y)) - (a   b))))$ $\mapsto$ $(0 >= y),(-(y) <= -(1))$  \\
\ttt{ps2-F.c}   & \rExact  & 29.0     & 1  & v,v & $\ell_{20}:!(((((c + (y   y)) - (2   x)) + y) < k))$ $\mapsto$ $(0 >= (-(c) + k)),((c - k) <= -(1))$  \\
\ttt{ps2-T.c}   & \rExact  & 53.4     & 2  & v,ep,v,v & $\ell_{20}:!(((((c + (y   y)) - (2   x)) + y) < k))$ $\mapsto$ $(((0 + (k   1)) + (y   -1)) <= 0),((-(k) + y) <= -(1))$  \\
\ttt{ps3-F.c}   & \rExact  & 62.8     & 2  & v,ep,v,v & $\ell_{20}:!((((((c + (6   x)) - (((2   y)   y)   y)) - ((3   y)   y)) - y) < k))$ $\mapsto$ $(((0 + (k   1)) + (c   -1)) <= 0),((c - k) <= -(1))$  \\
\ttt{ps3-T.c}   & \rExact  & 44.9     & 1  & v,v & $\ell_{20}:!((((((c + (6   x)) - (((2   y)   y)   y)) - ((3   y)   y)) - y) < k))$ $\mapsto$ $(0 >= (k - y)),((-(k) + y) <= -(1))$  \\
\ttt{ps4-F.c}   & \rExact  & 59.8     & 2  & v,ep,v,v & $\ell_{20}:!((((((c + (4   x)) - (((y   y)   y)   y)) - (((2   y)   y)   y)) - (y   y)) < k))$ $\mapsto$ $(((0 + (k   1)) + (y   -1)) <= 0),((-(k) + y) <= -(1))$  \\
\ttt{ps4-T.c}   & \rExact  & 58.1     & 2  & v,ep,v,v & $\ell_{20}:!((((((c + (4   x)) - (((y   y)   y)   y)) - (((2   y)   y)   y)) - (y   y)) < k))$ $\mapsto$ $(((0 + (k   1)) + (y   -1)) <= 0),((-(k) + y) <= -(1))$  \\
\ttt{ps5-F.c}   & \rExact  & 30.4     & 1  & v,v & $\ell_{19}:!(((((((c + (((((6   y)   y)   y)   y)   y)) + ((((15   y)   y)   y)   y)) + (((10   y)   y)   y)) - (30   x)) - y) < k))$ $\mapsto$ $(0 >= (-(c) + k)),((c - k) <= -(1))$  \\
\ttt{ps5-T.c}   & \rExact  & 41.3     & 1  & v,v & $\ell_{19}:!(((((((c + (((((6   y)   y)   y)   y)   y)) + ((((15   y)   y)   y)   y)) + (((10   y)   y)   y)) - (30   x)) - y) < k))$ $\mapsto$ $(0 >= (k - y)),((-(k) + y) <= -(1))$  \\
\ttt{ps6-F.c}   & \rExact  & 60.3     & 2  & v,ep,v,v & $\ell_{19}:\neg(c-2y^6 -6 y^5-5y^4+y^2+12x \leq k)$ $\mapsto$ $(((0 + (k   1)) + (c   -1)) <= 0),((c - k) <= -(1))$  \\
\ttt{ps6-T.c}   & \rExact  & 68.6     & 2  & v,ep,v,v & $\ell_{19}:\neg(c-2y^6 -6 y^5-5y^4+y^2+12x \leq k)$ $\mapsto$ $(((0 + (k   1)) + (y   -1)) <= 0),((-(k) + y) <= -(1))$  \\
\ttt{sqrt1-F.c} & \rExact  & 117.0    & 2  & v,en,v,v & $\ell_{21}:t^2-4s+2t+1+c \leq k$ $\mapsto$ $0 \geq a-k,k-a\leq -1$  \\
\ttt{sqrt1-T.c} & \rExact  & 116.3    & 2  & v,en,v,v & $\ell_{21}:t^2-4s+2t+1+c \leq k$ $\mapsto$ $0 \geq a-k,k-a\leq -1$  \\
\bottomrule

\end{tabular}}

\subsection{\tool{} on \benchpldi{} benchmarks}

{\footnotesize
\begin{tabular}{|l|c|r|r|p{0.8in}|p{2.8in}|}
\hline
\textbf{Benchmark} & \textbf{Res} & \textbf{Time} & \textbf{It.} &\textbf{Ref. Stages} &
\textbf{Output}  \\
\ttt{neg-afagp-F.c} & \rExact  & 289.1    & 2  & v,en,v,v & $\ell_{18}:z^2-12y-6z+12+c \leq k$ $\mapsto$ $0 \geq n-k,(((0 + (k   1)) + (n   -1)) <= -1)$  \\
\ttt{neg-afagp-T.c} & \rExact  & 287.2    & 2  & v,en,v,v & $\ell_{18}:z^2-12y-6z+12+c \leq k$ $\mapsto$ $0 \geq c-k,k-c \leq -1$  \\
\ttt{neg-afefp-F.c} & \rExact  & 225.2    & 2  & v,en,v,v & $\ell_{18}:z^2-12y-6z+12+c \leq k$ $\mapsto$ $0 \geq n-k,(((0 + (n   -1)) + (k   1)) <= -1)$  \\
\ttt{neg-afegp-F.c} & \rExact  & 226.0    & 2  & v,en,v,v & $\ell_{18}:z^2-12y-6z+12+c \leq k$ $\mapsto$ $0 \geq n-k,(((0 + (k   1)) + (n   -1)) <= -1)$  \\
\ttt{neg-afegp-T.c} & \rExact  & 217.4    & 2  & v,en,v,v & $\ell_{18}:z^2-12y-6z+12+c \leq k$ $\mapsto$ $0 \geq c-k,k-c \leq -1$  \\
\ttt{neg-afp-F.c} & \rExact  & 164.4    & 2  & v,en,v,v & $\ell_{18}:z^2-12y-6z+12+c \leq k$ $\mapsto$ $0 \geq n-k,(((0 + (k   1)) + (n   -1)) <= -1)$  \\
\ttt{neg-afp-T.c} & \rExact  & 160.9    & 2  & v,en,v,v & $\ell_{18}:z^2-12y-6z+12+c \leq k$ $\mapsto$ $0 \geq c-k,k-c \leq -1$  \\
\ttt{neg-efafp-F.c} & \rExact  & 243.0    & 2  & v,en,v,v & $\ell_{31}:z^2-12y-6z+12+c \leq k$ $\mapsto$ $0 \geq n-k,(((0 + (k   1)) + (n   -1)) <= -1)$  \\
\ttt{neg-egafp-F.c} & \rExact  & 282.2    & 2  & v,en,v,v & $\ell_{19}:z^2-12y-6z+12+c \leq k$ $\mapsto$ $0 \geq c-k,k-c \leq -1$  \\
\ttt{neg-egafp-T.c} & \rExact  & 313.3    & 2  & v,en,v,v & $\ell_{19}:z^2-12y-6z+12+c \leq k$ $\mapsto$ $0 \geq n-k,(((0 + (k   1)) + (n   -1)) <= -1)$  \\
\ttt{neg-egimpafp-T.c} & \rExact  & 251.4    & 2  & v,en,v,v & $\ell_{20}:z^2-12y-6z+12+c \leq k$ $\mapsto$ $0 \geq n-k,(((0 + (k   1)) + (n   -1)) <= -1)$  \\
\bottomrule

\ttt{afagp-F.c} & \rExact  & 103.9    & 2  & v,ep,v,v & $\ell_{18}:!(((((c + (y   y)) - (2   x)) + y) < k))$ $\mapsto$ $(((0 + (k   1)) + (c   -1)) <= 0),((c - k) <= -(1))$  \\
\ttt{afagp-T.c} & \rExact  & 74.8     & 1  & v,v & $\ell_{18}:!(((((c + (y   y)) - (2   x)) + y) < k))$ $\mapsto$ $(0 >= (-(c) + k)),((c - k) <= -(1))$  \\
\ttt{afefp-T.c} & \rExact  & 179.8    & 2  & v,en,v,v & $\ell_{19}:t^2-4s+2t+1+c \leq k$ $\mapsto$ $0 \geq a-k,k-a\leq -1$  \\
\ttt{afegp-F.c} & \rExact  & 239.7    & 2  & v,en,v,v & $\ell_{18}:t^2-4s+2t+1+c \leq k$ $\mapsto$ $0 \geq a-k,k-a\leq -1$  \\
\ttt{afegp-T.c} & \rExact  & 177.8    & 2  & v,en,v,v & $\ell_{19}:t^2-4s+2t+1+c \leq k$ $\mapsto$ $0 \geq c-k,k-c \leq -1$  \\
\ttt{afp-F.c}   & \rExact  & 163.8    & 2  & v,en,v,v & $\ell_{19}:z^2-12y-6z+12+c \leq k$ $\mapsto$ $0 \geq n-k,(((0 + (k   1)) + (n   -1)) <= -1)$  \\
\ttt{afp-T.c}   & \rExact  & 166.8    & 2  & v,en,v,v & $\ell_{19}:z^2-12y-6z+12+c \leq k$ $\mapsto$ $0 \geq c-k,k-c \leq -1$  \\
\ttt{agafp-F.c} & \rExact  & 65.7     & 1  & v,v & $\ell_{23}:z^2-12y-6z+12+c \leq k$ $\mapsto$ $0 \geq (c-k) \&\& (0 == (c-n)),((-(n) + k) <= -(1))$ \\
\ttt{agafp-T.c} & \rExact  & 36.3     & 1  & v,v & $\ell_{23}:z^2-12y-6z+12+c \leq k$ $\mapsto$ $0 \geq c-k,((-(c) + k) <= -(1))$  \\
\ttt{efafp-T.c} & \rExact  & 231.1    & 2  & v,en,v,v & $\ell_{29}:z^2-12y-6z+12+c \leq k$ $\mapsto$ $0 \geq n-k,(((0 + (k   1)) + (n   -1)) <= -1)$  \\
\bottomrule

\ttt{afagp-T.c} & \rExact  & 160.3    & 1  & v,v & $\ell_{21}:!(((((((x   z) - x) - y) + 1) + c) < k))$ $\mapsto$ $(0 >= (-(c) + k)),((c - k) <= -(1))$  \\
\ttt{afefp-T.c} & \rExact  & 312.1    & 2  & v,en,v,v & $\ell_{23}:((((((((2   Y)   xp) - ((2   X)   y)) - X) + (2   Y)) - v) + c) <= k)$ $\mapsto$ $0 \geq c-k,k-c \leq -1$  \\
\ttt{agafp-T.c} & \rExact  & 24.4     & 1  & v,v & $\ell_{22}:(((((c + (4   x)) - (((y   y)   y)   y)) - (((2   y)   y)   y)) - (y   y)) < k)$ $\mapsto$ $((c - k) <= -(1)),(0 >= (-(c) + k))$  \\
\ttt{neg-afefp-F.c} & \rExact  & 233.2    & 1  & v,v & $\ell_{17}:(((((1 + (x   z)) - x) - (z   y)) + c) < k)$ $\mapsto$ $((c - k) <= -(1)),(0 >= (-(c) + k))$  \\
\ttt{neg-afp-F.c} & \rExact  & 68.2     & 2  & v,en,v,v & $\ell_{17}:(((((c + (6   x)) - (((2   y)   y)   y)) - ((3   y)   y)) - y) < k)$ $\mapsto$ $((c - k) <= -(1)),(((0 + (k   1)) + (c   -1)) <= 0)$  \\
\bottomrule

\end{tabular}}

\subsection{\Tool{} on Handcrafted benchmarks}

{\scriptsize
\begin{tabular}{|l|c|r|r|p{0.35in}|p{3.2in}|}
\hline
\textbf{Benchmark} & \textbf{Res} & \textbf{Time} & \textbf{It.} &\textbf{Stages} &
\textbf{Output}  \\
\ttt{if-cubic-F.c} & \rExact  & 51.5     & 3  & v, tn, v, en, v, v & $\ell_{6}:(8 == ((x   x)   x))$ $\mapsto$ $(((4 >= (p + x)) \&\& (0 >= (p - x)) \&\& (0 >= (-(p) + x)) \&\& ((-(p) - x) <= -(4))) || ((2 >= p) \&\& (0 >= (-(p) + x)) \&\& ((-(p) - x) <= -(4)))), (((0 == (p - 2)) \&\& (1 >= (p - x)) \&\& !(((2 >= p) \&\& (0 >= (-(p) + x)) \&\& ((-(p) - x) <= -(4))))) || (0 >= x))$  \\
\ttt{if-cubic-T.c} & \rExact  & 51.3     & 3  & v, tn, v, en, v, v & $\ell_{6}:(8 == ((x   x)   x))$ $\mapsto$ $(((2 >= p) \&\& (0 >= (-(p) + x)) \&\& ((-(p) - x) <= -(4))) || ((-(x) <= -(2)) \&\& (4 >= (p + x)) \&\& (0 >= (-(p) + x)))), (((0 == (p - 2)) \&\& (0 >= -(x)) \&\& !(((-(x) <= -(2)) \&\& (4 >= (p + x)) \&\& (0 >= (-(p) + x))))) || (1 >= (p + x)))$  \\
\ttt{if-F.c}    & \rExact  & 78.8     & 6  & v, tn, v, ep, v, tp, v, tn, v, tp, v, v & $\ell_{6}:(36 == (x   x))$ $\mapsto$ $((((0 + (x   1)) <= 6) \&\& !(((0 == (p - 2)) \&\& ((-(p) + x) <= -(1))))) || ((x <= -(6)) \&\& (-(p) <= -(2)) \&\& (8 >= (p - x))))\&\&!(((0 == (p - 2)) \&\& (0 >= (p - x)) \&\& (3 >= (-(p) + x)))), (((((0 == (p - 2)) \&\& (0 >= -(x)) \&\& !(((2 >= p) \&\& (-(x) <= -(6)) \&\& (4 >= (-(p) + x))))) || ((0 == (p - 2)) \&\& ((-(p) + x) <= -(1)))) \&\& !(((x <= -(6)) \&\& (-(p) <= -(2)) \&\& (8 >= (p - x))))) || ((0 == (p - 2)) \&\& (0 >= (p - x)) \&\& (3 >= (-(p) + x))))$  \\
\ttt{if-T.c}    & \rExact  & 76.5     & 6  & v, tn, v, ep, v, tp, v, tn, v, tp, v, v & $\ell_{7}:(36 == (x   x))$ $\mapsto$ $(((((0 + (x   1)) + (p   1)) <= 8) \&\& !(((0 == (p - 2)) \&\& (3 >= x)))) || ((-(p) <= -(2)) \&\& (8 >= (p - x)) \&\& ((p + x) <= -(4))))\&\&!(((2 >= p) \&\& (5 >= x) \&\& (-(p) <= -(2)) \&\& ((-(p) - x) <= -(6)))), (((((0 == (p - 2)) \&\& (1 >= (p - x)) \&\& !(((6 >= x) \&\& (-(p) <= -(2)) \&\& ((p - x) <= -(4))))) || ((0 == (p - 2)) \&\& (3 >= x))) \&\& !(((-(p) <= -(2)) \&\& (8 >= (p - x)) \&\& ((p + x) <= -(4))))) || ((2 >= p) \&\& (5 >= x) \&\& (-(p) <= -(2)) \&\& ((-(p) - x) <= -(6))))$  \\
\ttt{square-loop-F.c} & \rExact  & 383.6    & 12 & v, tn, v, tn, v, tn, v, tn, v, tn, v, tn, v, tn, v, tn, v, tn, v, tn, v, tn, v, v & $\ell_{8}:(49 < (x   x))$ $\mapsto$ $((((((((((((-(x) <= -(8)) || ((0 == (p - 2)) \&\& (3 >= y) \&\& ((-(p) - y) <= -(3)) \&\& ((-(p) + x) <= -(10)))) || ((0 == (p - 2)) \&\& (7 >= (p + y)) \&\& ((x + y) <= -(4)) \&\& ((-(p) - y) <= -(6)))) || ((0 == (p - 2)) \&\& (10 >= y) \&\& ((-(p) - y) <= -(7)) \&\& ((-(p) + x) <= -(10)))) || ((0 == (p - 2)) \&\& (x <= -(8)) \&\& (10 >= (-(p) + y)) \&\& ((-(p) - y) <= -(13)))) || ((0 == (p - 2)) \&\& (x <= -(8)) \&\& (16 >= (-(p) + y)) \&\& ((-(p) - y) <= -(15)))) || ((0 == (p - 2)) \&\& ((p + x) <= -(6)) \&\& ((p - y) <= -(17)) \&\& (19 >= (-(p) + y)))) || ((0 == (p - 2)) \&\& (17 >= (x + y)) \&\& ((p - y) <= -(20)) \&\& ((-(p) + x) <= -(10)))) || ((0 == (p - 2)) \&\& ((p + x) <= -(6)) \&\& ((-(x) - y) <= -(18)))) || ((0 == (p - 1)) \&\& (14 >= y) \&\& (5 >= (x + y)) \&\& (0 >= (p - y)) \&\& ((-(p) + x) <= -(9)))) || ((0 == (p - 1)) \&\& (8 >= (x + y)) \&\& ((p - y) <= -(13)) \&\& ((-(p) + x) <= -(9)))) || ((0 == (p - 1)) \&\& ((-(x) - y) <= -(9)) \&\& ((-(p) + x) <= -(9)))), (7 >= x)\&\&!(((0 == (p - 2)) \&\& (3 >= y) \&\& ((-(p) - y) <= -(3)) \&\& ((-(p) + x) <= -(10))))\&\&!(((0 == (p - 2)) \&\& (7 >= (p + y)) \&\& ((x + y) <= -(4)) \&\& ((-(p) - y) <= -(6))))\&\&!(((0 == (p - 2)) \&\& (10 >= y) \&\& ((-(p) - y) <= -(7)) \&\& ((-(p) + x) <= -(10))))\&\&!(((0 == (p - 2)) \&\& (x <= -(8)) \&\& (10 >= (-(p) + y)) \&\& ((-(p) - y) <= -(13))))\&\&!(((0 == (p - 2)) \&\& (x <= -(8)) \&\& (16 >= (-(p) + y)) \&\& ((-(p) - y) <= -(15))))\&\&!(((0 == (p - 2)) \&\& ((p + x) <= -(6)) \&\& ((p - y) <= -(17)) \&\& (19 >= (-(p) + y))))\&\&!(((0 == (p - 2)) \&\& (17 >= (x + y)) \&\& ((p - y) <= -(20)) \&\& ((-(p) + x) <= -(10))))\&\&!(((0 == (p - 2)) \&\& ((p + x) <= -(6)) \&\& ((-(x) - y) <= -(18))))\&\&!(((0 == (p - 1)) \&\& (14 >= y) \&\& (5 >= (x + y)) \&\& (0 >= (p - y)) \&\& ((-(p) + x) <= -(9))))\&\&!(((0 == (p - 1)) \&\& (8 >= (x + y)) \&\& ((p - y) <= -(13)) \&\& ((-(p) + x) <= -(9))))\&\&!(((0 == (p - 1)) \&\& ((-(x) - y) <= -(9)) \&\& ((-(p) + x) <= -(9))))$  \\
\ttt{square-loop-T.c} & \rExact  & 233.3    & 7  & v, tn, v, tn, v, tn, v, tn, v, tn, v, tn, v, v & $\ell_{8}:(49 < (x   x))$ $\mapsto$ $(((((((-(x) <= -(8)) || ((0 == (p - 2)) \&\& (1 >= (p - y)) \&\& (5 >= (p + y)) \&\& ((p + x) <= -(6)))) || ((0 == (p - 2)) \&\& (5 >= y) \&\& (-(y) <= -(4)) \&\& ((x + y) <= -(4)))) || ((0 == (p - 2)) \&\& ((p + x) <= -(6)) \&\& ((x + y) <= -(1)) \&\& ((-(p) - y) <= -(7)))) || ((0 == (p - 2)) \&\& (0 >= (-(x) - y)) \&\& ((-(p) + x) <= -(10)))) || ((0 == (p - 1)) \&\& (0 >= (p - y)) \&\& ((x + y) <= -(6)) \&\& ((-(p) + x) <= -(9)))) || ((0 == (p - 1)) \&\& (x <= -(8)) \&\& (5 >= (-(x) - y)))), (7 >= x)\&\&!(((0 == (p - 2)) \&\& (1 >= (p - y)) \&\& (5 >= (p + y)) \&\& ((p + x) <= -(6))))\&\&!(((0 == (p - 2)) \&\& (5 >= y) \&\& (-(y) <= -(4)) \&\& ((x + y) <= -(4))))\&\&!(((0 == (p - 2)) \&\& ((p + x) <= -(6)) \&\& ((x + y) <= -(1)) \&\& ((-(p) - y) <= -(7))))\&\&!(((0 == (p - 2)) \&\& (0 >= (-(x) - y)) \&\& ((-(p) + x) <= -(10))))\&\&!(((0 == (p - 1)) \&\& (0 >= (p - y)) \&\& ((x + y) <= -(6)) \&\& ((-(p) + x) <= -(9))))\&\&!(((0 == (p - 1)) \&\& (x <= -(8)) \&\& (5 >= (-(x) - y))))$  \\
\end{tabular}}

(continued on next page)



\bigskip
{\scriptsize
\begin{tabular}{|l|c|r|r|p{0.35in}|p{3.2in}|}
\textbf{Benchmark} & \textbf{Res} & \textbf{Time} & \textbf{It.} &\textbf{Stages} &
\textbf{Output}  \\
\ttt{while-cubic-F.c} & \rExact  & 87.0     & 7  & v, tn, v, ep, v, tp, v, tn, v, tp, v, tn, v, v & $\ell_{6}:((64 >= ((y   y)   y)) \&\& (1 <= (y   y)))$ $\mapsto$ $(((((((0 + (p   1)) + (y   1)) <= 6) \&\& !(((0 >= y) \&\& (2 >= (p - y)) \&\& ((-(p) - y) <= -(2))))) || ((0 == (p - 1)) \&\& (5 >= (p + y)) \&\& ((p - y) <= -(1)))) \&\& !(((1 >= p) \&\& (0 >= -(y)) \&\& (-(p) <= -(1)) \&\& (4 >= (-(p) + y))))) || ((0 == (p - 1)) \&\& (4 >= y) \&\& ((p - y) <= -(1)))), (((((-(p) <= -(1)) \&\& (0 >= -(y)) \&\& ((-(p) - y) <= -(2)) \&\& !(((0 == (p - 2)) \&\& (4 >= y) \&\& ((-(p) - y) <= -(3))))) || ((0 >= y) \&\& (2 >= (p - y)) \&\& ((-(p) - y) <= -(2)))) \&\& !(((0 == (p - 1)) \&\& (5 >= (p + y)) \&\& ((p - y) <= -(1))))) || ((1 >= p) \&\& (0 >= -(y)) \&\& (-(p) <= -(1)) \&\& (4 >= (-(p) + y))))\&\&!(((0 == (p - 1)) \&\& (4 >= y) \&\& ((p - y) <= -(1))))$  \\
\ttt{while-cubic-T.c} & \rExact  & 84.4     & 7  & v, tn, v, ep, v, tp, v, tn, v, tp, v, tn, v, v & $\ell_{6}:((64 >= ((y   y)   y)) \&\& (1 <= (y   y)))$ $\mapsto$ $(((((((0 + (p   1)) + (y   1)) <= 6) \&\& !(((0 >= -(y)) \&\& (-(p) <= -(2)) \&\& (2 >= (p + y))))) || ((0 == (p - 1)) \&\& (4 >= y) \&\& (-(y) <= -(2)))) \&\& !(((1 >= p) \&\& (5 >= y) \&\& (0 >= -(y)) \&\& (-(p) <= -(1))))) || ((0 == (p - 1)) \&\& ((p - y) <= -(1)) \&\& (3 >= (-(p) + y)))), (((((0 >= -(y)) \&\& ((-(p) - y) <= -(2)) \&\& !(((0 == (p - 2)) \&\& (4 >= y) \&\& (-(y) <= -(1))))) || ((0 >= -(y)) \&\& (-(p) <= -(2)) \&\& (2 >= (p + y)))) \&\& !(((0 == (p - 1)) \&\& (4 >= y) \&\& (-(y) <= -(2))))) || ((1 >= p) \&\& (5 >= y) \&\& (0 >= -(y)) \&\& (-(p) <= -(1))))\&\&!(((0 == (p - 1)) \&\& ((p - y) <= -(1)) \&\& (3 >= (-(p) + y))))$  \\
\ttt{while-F.c} & \rExact  & 190.4    & 6  & v, tn, v, tn, v, tn, v, tn, v, tn, v, v & $\ell_{9}:(63 <= ((x   x) - (2   x)))$ $\mapsto$ $((((((((p - x) <= -(7)) \&\& ((-(p) - x) <= -(10))) || ((0 == (p - 2)) \&\& (1 >= (p - y)) \&\& (0 >= (-(p) + y)) \&\& ((x + y) <= -(6)))) || ((0 == (p - 2)) \&\& (0 >= (p - y)) \&\& ((x + y) <= -(5)))) || ((0 == (p - 2)) \&\& (4 >= (-(x) - y)) \&\& ((-(p) + x) <= -(9)))) || ((0 == (p - 1)) \&\& (4 >= (p + y)) \&\& (0 >= (p - y)) \&\& ((x + y) <= -(8)))) || ((0 == (p - 1)) \&\& (0 >= (p - y)) \&\& ((-(p) + x) <= -(8)))), (8 >= x)\&\&!(((0 == (p - 2)) \&\& (1 >= (p - y)) \&\& (0 >= (-(p) + y)) \&\& ((x + y) <= -(6))))\&\&!(((0 == (p - 2)) \&\& (0 >= (p - y)) \&\& ((x + y) <= -(5))))\&\&!(((0 == (p - 2)) \&\& (4 >= (-(x) - y)) \&\& ((-(p) + x) <= -(9))))\&\&!(((0 == (p - 1)) \&\& (4 >= (p + y)) \&\& (0 >= (p - y)) \&\& ((x + y) <= -(8))))\&\&!(((0 == (p - 1)) \&\& (0 >= (p - y)) \&\& ((-(p) + x) <= -(8))))$  \\
\ttt{while-T.c} & \rExact  & 189.6    & 6  & v, tn, v, tn, v, tn, v, tn, v, tn, v, v & $\ell_{9}:(63 <= ((x   x) - (2   x)))$ $\mapsto$ $((((((-(x) <= -(9)) || ((0 == (p - 2)) \&\& (1 >= (p - y)) \&\& ((x + y) <= -(6)) \&\& (0 >= (-(p) + y)))) || ((0 == (p - 2)) \&\& (x <= -(7)) \&\& (0 >= (p - y)))) || ((0 == (p - 1)) \&\& (0 >= (p - y)) \&\& ((x - y) <= -(9)) \&\& (2 >= (-(p) + y)) \&\& ((x + y) <= -(5)))) || ((0 == (p - 1)) \&\& (x <= -(7)) \&\& (0 >= (p - y)) \&\& ((x + y) <= -(2)))) || ((0 == (p - 1)) \&\& (1 >= (-(x) - y)) \&\& ((-(p) + x) <= -(8)))), (8 >= x)\&\&!(((0 == (p - 2)) \&\& (1 >= (p - y)) \&\& ((x + y) <= -(6)) \&\& (0 >= (-(p) + y))))\&\&!(((0 == (p - 2)) \&\& (x <= -(7)) \&\& (0 >= (p - y))))\&\&!(((0 == (p - 1)) \&\& (0 >= (p - y)) \&\& ((x - y) <= -(9)) \&\& (2 >= (-(p) + y)) \&\& ((x + y) <= -(5))))\&\&!(((0 == (p - 1)) \&\& (x <= -(7)) \&\& (0 >= (p - y)) \&\& ((x + y) <= -(2))))\&\&!(((0 == (p - 1)) \&\& (1 >= (-(x) - y)) \&\& ((-(p) + x) <= -(8))))$  \\
\bottomrule

\end{tabular}}

\vfill

\end{document}